\documentclass[a4paper,usenatbib]{mnras}
\usepackage{newtxtext,newtxmath}

\usepackage[T1]{fontenc}
\usepackage{ae,aecompl}

\usepackage[utf8]{inputenc}
\usepackage{geometry}
\usepackage{graphicx}
\usepackage{epstopdf}
\usepackage{epsfig}

\usepackage{hyperref}
\usepackage{color}
\usepackage{amsmath}

\usepackage{amssymb}	% Extra maths symbols

\usepackage{verbatim}

\title[A low-luminosity CC supernova]{A low-luminosity core collapse supernova very similar to SN 2005cs}

\author[J칛ger et al.]{Zolt치n  J칛ger Jr.$^{1,2}$,
J칩zsef  Vink칩$^{2,3,5}$,
Barna I. B\'ir칩$^{1}$,
Tibor Heged\"us$^{1}$, %콤
\newauthor
Tam치s  Borkovits$^{1,3}$,
Zolt치n  J칛ger Sr.$^{1}$,
Andrea  P. Nagy$^{2,3}$,
L치szl칩  Moln치r$^{3,4,5}$,
\newauthor
Levente  Kriskovics$^{3,5}$
\\
\\
\\
$^{1}$Baja Astronomical Observatory of the University of Szeged, Szegedi 칰t KT766, H-6500 Baja, Hungary\\
$^{2}$Department of Optics and Quantum Electronics, University of Szeged, D칩m t칠r 9, H-6720 Szeged, Hungary\\
$^{3}$Konkoly Observatory, Research Center for Astronomy and Earth Sciences, Konkoly Thege Mikl칩s 칰t 15-17., Budapest H-1121, Hungary\\ %Research Centre for Astronomy and Earth Sciences
$^{4}$CSFK Lend\"ulet Near-Field Cosmology Research Group, Konkoly Thege 15-17, H-1121 Budapest, Hungary\\
$^{5}$ELTE E\"otv\"os Lor\'and University, Institute of Physics, P\'azm\'any P\'eter s\'et\'any 1/A, Budapest, 1117 Hungary
}

\date{Accepted 2020 June 9. Received 2020 June 9; in original form 2019 September 24}

\pubyear{2020}

\begin{document}
\label{firstpage}
\pagerange{\pageref{firstpage}--\pageref{lastpage}}
\maketitle

\begin{abstract}

   \noindent We present observations and analysis of PSN J17292918+7542390, 
   a low-luminosity Type II-P supernova (LL SN IIP). The observed sample of such events is still low, and their nature is still under debate.
   Such supernovae are similar to
   SN 2005cs, a well-observed low-luminosity Type II-P event, having low expansion velocities, and small ejected $^{56}$Ni mass.   
   We have developed a robust and relatively fast Monte-Carlo code that fits semi-analytic models to 
   light curves of core collapse supernovae. 
   This allows the estimation of the most important physical parameters, like the radius of the progenitor star, 
   the mass of the ejected envelope, the mass of the radioactive nickel synthesized during the explosion, among others.   
   PSN J17292918+7542390 has $R_0 = 91_{-70}^{+119} \cdot 10^{11} \;\text{cm}$, $M_\text{ej} = 9.89_{-1.00}^{+2.10} \; M_{\odot}$, 
   $E_{\mbox{kin}} = 0.65_{-0.18}^{+0.19} \;\text{foe}$, $v_{\mbox{exp}} = 3332_{-347}^{+216}$ km s$^{-1}$,
   for its progenitor radius, ejecta mass, kinetic energy and expansion velocity, respectively.
   The initial nickel mass of the PSN J17292918+7542390 turned out to be $1.55_{-0.70}^{+0.75} \cdot 10^{-3} M_{\odot}$. 
   The measured photospheric velocity at the earliest observed phase is 7000 km s$^{-1}$.
   As far as we can tell based on the small population of observed low-luminosity Type II-P supernovae, the determined values are typical for these events.

\end{abstract}

%km/s
%maximum 6!

%\tableofcontents

\section{Introduction}

%Az \'ekezetes bet\H uk kiszed\'es\'ere a r\'egi TeX-ez\H ok szinte mind az ilyen id\'etlen jel\"ol\'st haszn\'alj\'ak.
%{\bf or black hole (\citealt{Burrows-SNtheory-2})}.

Core-collapse supernovae (CC SNe) are thought to originate from stars having $M_{ZAMS} > 8$~$M_\odot$, 
following the collapse of their iron core to a neutron star
or black hole (\citealt{Burrows-SNtheory}; \citealt{Burrows-SNtheory-2}).
Stars that retain their massive hydrogen-rich envelope produce Type II-P and II-L SNe, 
showing strong H features in their spectra and a plateau (with a wide range in duration) in their optical light curves.%$\sim$100 day-long
In the past 20 years the existence of a subclass of low-luminosity (LL) Type II-P SNe has been recognized 
(\citealt{Pastorello-sublum}; \citealt{spiro-sublum}). 
These events have absolute magnitudes fainter by $\sim$2--3 mag than regular Type II-P SNe 
(absolute magnitude between -16 -- -19 mag),
have lower expansion velocities, and seem to produce less amount of radioactive nickel during explosion.

The prototype of this group was SN~1997D (\citealt{Turatto-ecapture}), which unfortunately was discovered at later phase 
giving high uncertainties in the explosion epoch (\citealt{benetti-1997d}).
The origin of these events is still under debate.

They may arise from a moderate mass ($\sim$9$ M_{\odot}$) star, or 
from a massive star ($\sim$25$ M_{\odot}$) that failed to explode entirely (\citealt{Turatto-ecapture}; 
\citealt{Chugai-ecapture}; \citealt{Kitaura-ecapture}). 
Light-curve modeling and progenitor identification from pre-explosion images 
(e.g. for 2005cs: $M_0 = 6-13 \; M_{\odot}$ by \citealt{Maund-05csrefHST}; \citealt{Li-05csrefHST}; \citealt{Eldridge-05csrefHST})
suggest they are likely to arise from moderate mass red supergiant (RSG) stars 
(\citealt{Pumo-sublum}; \citealt{Sergey-sublum} and references therein) with masses between 8--12 $M_{\odot}$.
However the massive star scenario may also be possible: SN~2016bkv had an ejected mass between 16--19 $M_{\odot}$ (\citealt{tatsuya-massiveLL}).
They might also be electron capture SNe, where, instead of a collapsing Fe core, 
an O-Ne-Mg core captures the electrons, which leads to a subsequent collapse
(see \citealt{Kitaura-ecapture}; \citealt{Wanajo-ecapture}).
However iron core collapse is also a possibility: for example, SN~1997D had more likely an iron core rather than O-Ne-Mg core 
(\citealt{Jerkstrand-EC/Fecore}). %late spectral modelling
The sample of sub-luminous SNe is still small, they represent only 5\% of all the Type II SNe events (\citealt{Pastorello-sublum}), 
so any new observed event may be important to investigate. 

Assuming a Salpeter initial mass function (IMF) with an exponent of 2.3, 42\% of the massive stars are born with masses between 8--12 $M_{\odot}$,
and 15\% in the range 8--9 $M_{\odot}$. The range may be even narrower for these low-luminous events, 
but we probably miss many of them due to their fainter absolute brightness. \\

PSN J17292918+7542390, hereafter SN-NGC6412, a LL Type II-P SN, was discovered on 2015-07-10 (57213 MJD) by Ron Arbour amateur astronomer (\citealt{toma-spectra}).  % , \citealt{toma-spectra}).
A single spectrum was made one day after the discovery epoch (57214 MJD) by \citet{toma-spectra}.  %discovery is tomascellaban van benne
No additional data have been published for this event.

In this paper we present photometric observations of SN-NGC6412 taken at two observatories in Hungary. This object turned out to be
a low luminosity IIP SN with low velocities, and small ejected $^{56}$Ni mass.
Even though our follow-up observations missed the end of the plateau, 
the rarity of such events may still make the data and the analysis valuable.\\

This paper also contains the modeling of the bolometric light curve (LC) of SN-NGC6412 using a semi-analytic model. 
Such simple light curve (LC) models have been developed as early as the 1980's by \citet{Arnett-LC1, Arnett-LC2}  %\citet{Arnett-LC1}; \citet{Arnett-LC2} 
to explain and fit bolometric light curves of supernovae. 
Later they were further improved by the inclusion of numerical computations of certain aspects, 
thus becoming semi-analytic (\citealt{Arnett-Fu-LC}; \citealt{Popov-LC}). 
These models assume spherical symmetry and do not adequately model the initial transient behavior at early stages (t<20 days), 
which obviously makes them inferior compared to detailed hydrodynamical models; nevertheless, 
they are useful to derive estimates or constraints of basic parameters like the explosion energy, 
ejected mass and initial radius, without the high computational resources demanded by the latter. 

Applications of the code so far indicate that it can model IIP LC particularly well,
while other SN types may have further physical mechanisms which is not considered in this model:
Circumstellar medium (CSM) interaction for IIn (narrow emission line in spectra) and IIL (no significant plateau present), 
and binary companion for IIb, Ib and Ic (weak or no H and silicon present in spectra) (see \citealt{Nagy-Ibc}).
They can also give first-order approximations in cases of limited observational information, 
for example when only photometric time series are available for a particular SN. 
Fitting the parameters of such models to the light curve has been traditionally made "by hand" using a trial and error approach. 
Strong correlations between the physical parameters, however, make this procedure complicated (\citealt{Arnett-Fu-LC}; \citealt{Nagy-LC}). 
This approach also lacks a firm estimate of the parameter uncertainties. 
However, in the age of high computing powers, a method of Markov Chain Monte Carlo (MCMC) sampling becomes 
feasible (\citealt{Metropolis-mcmc}; \citealt{Hastings-mcmc}). 
MCMC is a well-established technique for constraining parameters from observed data, and especially suited for cases 
when the parameter space has a high dimensionality (\citealt{Gilks-mcmc}). 
The sampling maps the whole parameter space based on the joint posterior probability distribution of all the parameters. 
It has the  nice ergodic property which allows the various integrals over the parameter space (mean
value, standard deviation, and percentiles in particular) to be computed as simple sums over the chain elements. 
Therefore it makes possible to provide not only best estimates of the parameters, 
but also confidence intervals for them (leading to uncertainty determinations); furthermore, 
correlations between the various parameters can also be revealed.

In Section 2 we describe the details of observations of SN-NGC6412, 
and then we make an approximation for the dust extinction in Section 3.
In Section 4 we compare the light curve of the SN-NGC6412 with other SNe. Section 5 presents the analysis of SN-NGC6412 spectrum. 
In Section 6 we fit diluted blackbody radiation to the photometric filters to determine the temperatures and photometric radii at various times of observation.
Sections 7 -- 9 present the bolometric light curve calculation, and the MCMC analysis of SN-NGC6412.
In Section 10 we present the results, and finally Section 11 summarizes the main conclusions of this paper.
The Appendix contains a technical description of the MCMC fitting program used in the analysis.

\section{Photometry}

For the photometric monitoring of SN-NGC6412 we used two optical telescopes.
One of them is a 50~cm telescope of Baja Observatory, equipped with SDSS griz filters.
The other is the 60/90~cm Schmidt telescope located at Piszk\'{e}stet\H{o} Mountain Station of Konkoly Observatory, and uses Johnson BVRI filters.
Our photometric monitoring of SN-NGC6412 began on 2015-07-12, 2 days after its discovery,
and continued for $\sim 100$ days up to the end of the plateau phase. 
After emerging from solar conjunction, the last observation was made
after 250 days with both telescopes, however, by that time the SN became very dim.

The data were reduced with standard prescription in IRAF environment. We used Point Spread Function (PSF) photometry to obtain the magnitude values. 
Nearby fields covered by the SDSS survey were observed during the same nights 
as the SN and the SDSS DR15 catalog\footnote{ http://skyserver.sdss.org/dr9/en/tools/chart/navi.asp} \citep{dr13}
was used to determine standard magnitudes of stars in the target field. %\citealt{sdss-dr9} 
(The catalog unfortunately does not cover the field of NGC~6412 containing the supernova.)
These magnitudes were then used to tie the magnitudes of the SN to standard systems. 
%The uncertainty comes from the standard deviation of the difference in measured and reference data of the standard comparison stars.
The uncertainties were computed using standard propagation of errors, the major contribution coming from the standard calibration.
Because of the two different filter systems, one of them must be converted to the other using empirical relations measured by \citet{jordi-filter}. 
We chose to convert the SDSS~griz to Johnson~BVRI. 
The calibrated SN magnitudes are reported in Table ~\ref{phottabl}.
The SN appeared in a brighter region (possibly a \ion{H}{II} region) which has non-negligible contribution to the measured brightness of the SN.
This region was observed $\sim$2 years after the peak (MJD = 57834.5) when the SN already dimmed below the detection limit, 
and its fluxes were  %($m_\text{ref}$) 
subtracted from the SN fluxes. 

The last epoch when the SN was detected (MJD = 57464.5) was $\sim$254 days after explosion (see Section 6 for a discussion of the date of explosion ($t_0$)). 
At this phase the SN was only slightly (0.2--0.3 mag) brighter than the underlying \ion{H}{II} region in all bands, 
which makes the uncertainty of the measured brightnesses quite high. 
Nevertheless, the faintness of SN-NGC6412 at such a late phase suggests low amounts of initial nickel mass.

\begin{figure*}
%\begin{center}
\includegraphics[width=\textwidth]{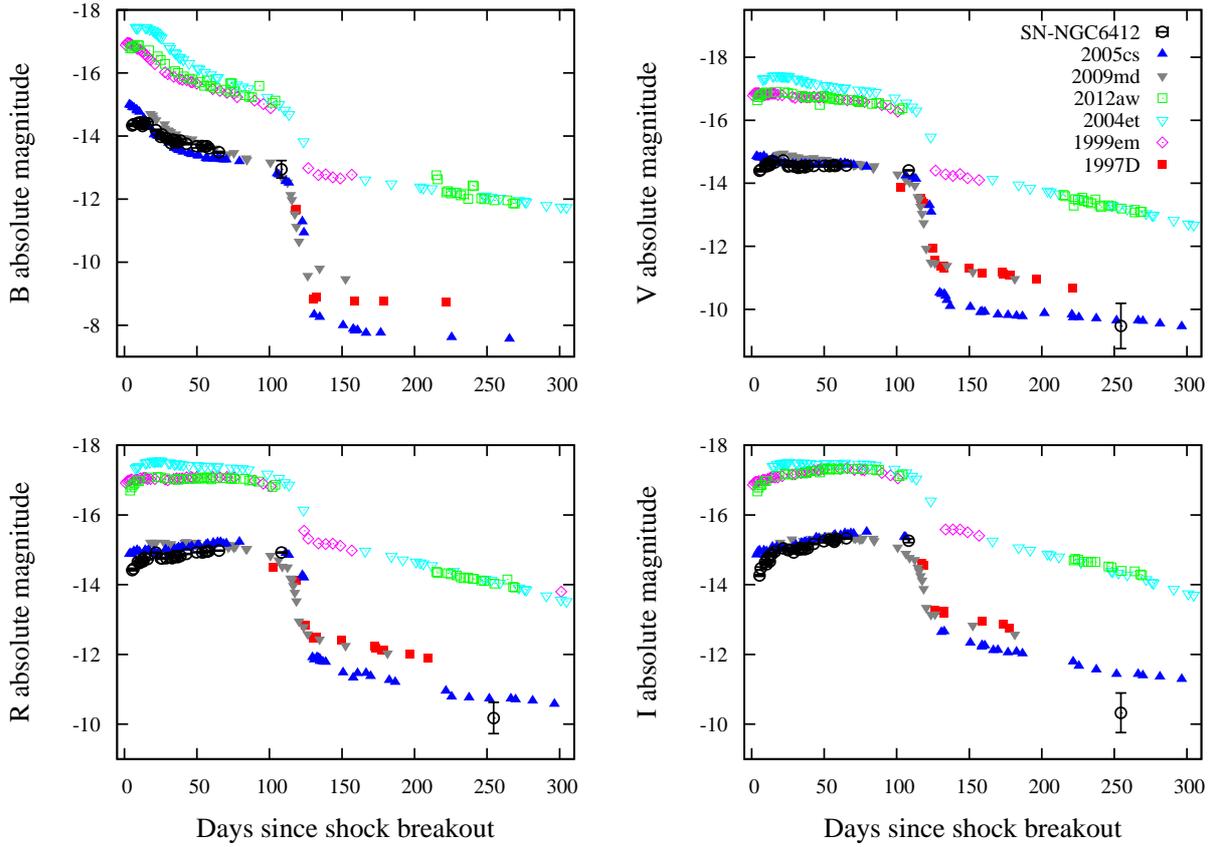} 
%\framebox[6in][c]{\raisebox{0pt}[2in][2in]{Fig1}}
\caption{De-reddened absolute BVRI light curves of SN-NGC6412 together with those of other Type II-P SNe. 
Parameters for calculating the absolute light curves are taken from Table ~\ref{Adopted}.
The similarity between SN-NGC6412 and the low-luminosity Type II-P events (SN 1997D, 2005cs, 2009md) is apparent. 
Note: most of the time, the error bars are smaller than the point size.
%Explosion epoch discuss for SN-NGC6412 is can be found in Section 6.
}
\label{Phothas}
%\end{center}
\end{figure*}

Fig.~\ref{Phothas} shows the LC of SN-NGC6412 in BVRI-bands, converted to absolute magnitudes (see Section 4), 
and compared to those of other Type II-P SNe. 
The presence of the plateau as well as the low absolute magnitudes make SN-NGC6412 a member of the LL Type II-P SN subclass. 

%\begin{equation}
%m' = m - 2.5 \cdotp log_{10}( 1 - 10^{-0.4 \cdotp (m_{ref} - m) } )  
%\end{equation}

\begin{table*}
\begin{center}
\caption{BVRI photometry of SN-NGC6412. All data are transformed to the Johnson-Cousins BVRI system. 
No further corrections have been applied. 
The data taken on the last epoch (at +625d) were used to correct for the contamination from the underlying \ion{H}{II} region. }
\label{phottabl}
\begin{tabular}{ l c c c c c c c c c c }
\hline

{\bf MJD} & {\bf Phase} & {\bf B} & {\bf B err}  & {\bf V} & {\bf V err}  & {\bf R} & {\bf R err}  & {\bf I} & {\bf I err}  & {\bf Inst.} \\
\hline
57215.5  &  5.5   & 16.948   & 0.027   & 16.863   & 0.029   & 16.792   & 0.012   & 16.833   & 0.012   &  1 \\
57216.3  &   6.3  &  16.916  &  0.025  &  16.832  &  0.024  &  16.755  &  0.012  &  16.665  &  0.032  &  1 \\
57219.3  &   9.3  &  16.876  &  0.031  &  16.708  &  0.022  &  16.578  &  0.012  &  16.517  &  0.026  &  1 \\
57220.5  &  10.5  &  16.895  &  0.018  &  16.734  &  0.015  &  16.611  &  0.006  &  16.565  &  0.009  &  1 \\
57221.0  &  11.0  &  16.862  &  0.049  &  16.649  &  0.020  &  16.574  &  0.018  &  16.412  &  0.038  &  2 \\
57222.5  &  12.5  &  16.966  &  0.036  &  16.705  &  0.017  &  16.520  &  0.008  &  16.497  &  0.017  &  1 \\
57223.0  &  13.0  &  16.911  &  0.039  &  16.667  &  0.017  &  16.566  &  0.018  &  16.336  &  0.029  &  2 \\
57224.0  &  14.0  &  16.872  &  0.033  &  16.583  &  0.013  &  16.471  &  0.011  &  16.309  &  0.024  &  2 \\
57226.0  &  16.0  &  16.887  &  0.051  &  16.624  &  0.020  &  16.509  &  0.016  &  16.250  &  0.037  &  2 \\
57231.8  &  21.8  &  17.080  &  0.025  &  16.564  &  0.011  &  16.338  &  0.012  &  16.163  &  0.028  &  2 \\
57235.0  &  25.0  &  17.268  &  0.080  &  16.737  &  0.027  &  16.487  &  0.011  &  16.210  &  0.032  &  2 \\
57236.0  &  26.0  &  17.200  &  0.050  &  16.722  &  0.019  &  16.495  &  0.016  &  16.201  &  0.041  &  2 \\
57239.0  &  29.0  &  17.299  &  0.019  &  16.740  &  0.010  &  16.486  &  0.012  &  16.272  &  0.023  &  2 \\
57242.0  &  32.0  &  17.434  &  0.024  &  16.778  &  0.010  &  16.466  &  0.008  &  16.196  &  0.016  &  2 \\
57243.0  &  33.0  &  17.343  &  0.027  &  16.767  &  0.011  &  16.487  &  0.009  &  16.162  &  0.016  &  2 \\
57244.0  &  34.0  &  17.345  &  0.026  &  16.737  &  0.010  &  16.452  &  0.007  &  16.192  &  0.013  &  2 \\
57244.8  &  34.8  &  17.425  &  0.054  &  16.748  &  0.020  &  16.427  &  0.014  &  16.158  &  0.023  &  2 \\
57245.8  &  35.8  &  17.366  &  0.024  &  16.735  &  0.015  &  16.436  &  0.020  &  16.162  &  0.039  &  2 \\
57246.8  &  36.8  &  17.418  &  0.038  &  16.765  &  0.015  &  16.455  &  0.012  &  16.187  &  0.026  &  2 \\
57248.8  &  38.8  &  17.364  &  0.031  &  16.727  &  0.012  &  16.425  &  0.011  &  16.160  &  0.037  &  2 \\
57252.8  &  42.8  &  17.451  &  0.016  &  16.726  &  0.008  &  16.372  &  0.009  &  16.047  &  0.019  &  2 \\
57259.8  &  49.8  &  17.513  &  0.055  &  16.722  &  0.018  &  16.330  &  0.009  &  15.978  &  0.048  &  2 \\
57263.8  &  53.8  &  17.510  &  0.032  &  16.729  &  0.016  &  16.334  &  0.018  &  15.933  &  0.039  &  2 \\
57265.8  &  55.8  &  17.520  &  0.038  &  16.716  &  0.017  &  16.316  &  0.017  &  15.951  &  0.033  &  2 \\
57267.0  &  57.0  &  17.513  &  0.043  &  16.704  &  0.016  &  16.310  &  0.010  &  15.994  &  0.023  &  2 \\
57267.3  &  57.3  &  No data & No data &  16.723  &  0.011  &  16.307  &  0.014  &  15.946  &  0.020  &  1 \\
57268.3  &  58.3  &  17.570  &  0.014  &  16.721  &  0.007  &  16.295  &  0.011  &  15.915  &  0.025  &  1 \\
57275.3  &  65.3  &  17.670  &  0.012  &  16.716  &  0.009  &  16.286  &  0.006  &  15.900  &  0.022  &  1 \\
57318.3  & 108.3  &  18.073  &  0.272  &  16.852  &  0.008  &  16.334  &  0.014  &  15.966  &  0.053  &  1 \\
57464.5  & 254.5  &  19.393  &  0.236  &  18.877  &  0.134  &  No data & No data &  No data & No data &  2 \\
57464.5  & 254.5  &  19.335  &  0.133  &  19.141  &  0.127  &  18.606  &  0.090  &  18.222  &  0.097  &  1 \\
57834.5  & 624.5  &  19.375  &  0.092  &  19.404  &  0.104  &  19.087  &  0.080  &  18.656  &  0.098  &  1 \\

\hline
\end{tabular}
\end{center}
$^{1}$ 60/90 cm Schmidt telescope, BVRI filters, Konkoly Observatory, Piszkestet콈, Hungary\\
$^{2}$ 50cm RC-telescope, SDSS griz filters, Baja Observatory, Baja, Hungary
%\caption{SN-NGC6412 observed values}

\end{table*}

\section{Estimation of dust extinction}

The interstellar reddening due to the Milky Way dust in the direction of SN-NGC6412 is $E(B-V) = 0.035$ mag acquired from the 
NASA Extragalactic Database (NED; \citealt{Schlafly-ref-NED}). %\citealt{ref-NED}

The total extinction that includes the effect of intergalactic dust as well as the dust within the host galaxy 
can be estimated from the equivalent width (EW) of the Na~D line, 
which correlates with $E(B-V)$ (see \citealt{poznanski-extinction} for more information). 
Combining the unresolved Na D$_1$ and D$_2$ features one can get 
\begin{equation}
\log_{10} E(B-V)_{} = 1.17 \cdotp \mbox{EW} (\mbox{D}_1+\mbox{D}_2) - 1.85 \pm 0.08  
\end{equation}
%\begin{equation}
%A_V = 3.1 \cdotp E(B-V)
%\end{equation}
%E(B-V): 0.0212
%ned: $A_V = 0.11$
%ned: E(B-V): 0.035

Examining the observed spectrum of SN-NGC6412 (see Section 5) 
around the Na D line (Fig.~\ref{NaD} top right panel), shows no features exceeding the noise level significantly.  
The panel shows the Na D line of the host and the milky way in the spectra. They are not in the same position due the redshift.
The weak features appearing in the spectrum are probably noise, implying low host extinction, 
 which is in agreement with the fact that the galaxy is seen face-on.
Therefore we neglect the intergalactic and host galaxy extinction, and adopt $E(B-V)= 0.035$ mag.

%\begin{figure}
%\begin{center}
%\includegraphics[width=\columnwidth]{eps/NaDill.eps} 
%\framebox[6in][c]{\raisebox{0pt}[2in][2in]{Fig1}}
%\caption{The Na D line in the SN-NGC6412 spectrum.}
%\label{NaD}
%\end{center}
%\end{figure}

\section{Comparison with SN~2005cs and other SNe}

A comparison of the LCs of SN-NGC6412 and SN~2005cs (\citealt{dessart-2005cs}; \citealt{Pastorello2006}; \citealt{pastorello-2005cs}) 
reveals some remarkable similarities, as shown in Fig.~\ref{Phothas}, 
and also in Table ~\ref{Adopted} of the adopted values.

Despite its relative proximity, the distance to the host galaxy, NGC~6412, is quite uncertain. 
The NASA Extragalactic Database\footnote{https://ned.ipac.caltech.edu} (NED) lists several values determined from various methods. %(\citealt{ref-NED})
\citet{Bottinelli-distance, Bottinelli-distance2}  %\citet{Bottinelli-distance}, \citet{Bottinelli-distance2} 
acquired distances between 12.4 and 14.9 Mpc with the Tully-Fisher (T-F) method, 
but with quite high ($\sim 4$-5 Mpc) uncertainty. They assumed $H_0 = 103$ km~s$^{-1}$~Mpc$^{-1}$ for the value of the Hubble-constant.
On the other hand, \citet{Tully-Fisher} obtained $23.5 \pm 4.8$ Mpc from the same method, assuming $H_0 = 75$ km~s$^{-1}$~Mpc$^{-1}$. 
To our knowledge, no other redshift independent distance estimate is available for NGC~6412.

The issue with this inhomogeneous dataset is twofold. 
First, they are tied to different calibrations relating the relative distances and the absolute scale, 
reflected by the different values of the Hubble-constant. 
Second, NGC~6412 is a face-on galaxy, which renders the measured rotation velocities uncertain.

Using a more recently determined Hubble-constant may reduce the systematic offset between the various calibrations. 
At present there is a well-known tension between the $H_0$ parameters measured either in the local Universe or at high redshifts: 
from CMB fluctuations $H_0 = 67.37 \pm 0.54$ km~s$^{-1}$~Mpc$^{-1}$ was obtained by the Planck mission (\citealt{plank-H0}), 
while Cepheids and Type Ia supernovae in the local Universe resulted in $H_0 = 73.48 \pm 1.66$ km~s$^{-1}$~Mpc$^{-1}$ (\citealt{riess-H0}). 
If we adopt the mean of these two measurements, $H_0 \sim 71$ km~s$^{-1}$~Mpc$^{-1}$, 
then the T-F distances by \citet{Bottinelli-distance, Bottinelli-distance2} will increase to values between 18 and 21.6 Mpc, 
while the distance given by \citet{Tully-Fisher} will be $\sim 24.8$ Mpc, thus, 
reducing the gap between the various distance measurement results.

NED also gives redshift-dependent distances based on the kinematics of NGC~6412.
The recession velocity corrected for the infall to the Virgo cluster (\citealt{mould-virgoinfall}) is $v_{Virgo} = 1715 \pm 18$ km~s$^{-1}$, 
which gives $D_{kin} \sim 24.1 \pm 0.25$ Mpc adopting $H_0 = 71$ km~s$^{-1}$~Mpc$^{-1}$ as above. 
On the other hand, the velocity with respect to the 3K Cosmic Background Radiation (CMB), 
$v_{CMB} = 1262 \pm 4 $km~s$^{-1}$ (\citealt{fixsen-CMB}) would result in $D_{CMB} \sim 17.8 \pm 1.3$ Mpc. 
These estimates have about the same amount of discrepancy ($\sim 6$ Mpc) as the results from the T-F method listed above.

Unfortunately, the various distance measurement methods that use the SN itself cannot be applied in this case. 
The Expanding Photosphere Method (EPM, \citealt{Kirshner-EPM}) needs velocity information, 
which is not available here, because of the single spectrum of SN-NGC6412 observed at very early phase. 
The application of that spectrum would give a very high ($\sim 20$ Mpc) uncertainty for the EPM-distance. 
The Standard Candle Method for Type II-P SNe (\citealt{hamuy-SCM}) is also less reliable for LL SNe II.

After finding no additional constraint, we decided to adopt the $D_{CMB} \sim 17.8$ Mpc distance, 
that is also close to the rescaled T-F result from \citet{Bottinelli-distance}. 
The motivation for choosing this value is that in this case the absolute light curves of SN-NGC6412 are overlap with those of SN~2005cs. 
While this is not necessarily true, as LL SNe II display a range of absolute magnitudes, 
this is the only additional constraint that can be applied in the distance estimate. 
If the longer, $D \sim 24$ Mpc distance were the true one, then SN-NGC6412 would be $\sim 0.6$ mag brighter, 
which must be kept in mind while comparing the absolute magnitudes to those of other SNe II-P.

Because of the remarkable similarities, %(\citealt{pastorello-2005cs}), 
we used the very well-observed SN~2005cs as the main reference for comparative analysis.
SN 2005cs was also a sub-luminous, $^{56}$Ni-poor, low-energy Type II-P SN (\citealt{pastorello-2005cs}) similar to SN-NGC6412.
While the B band fluxes evolved slightly differently, the V, R and I band light curves look very similar.
Although the brightnesses at the earliest phases are a bit fainter for SN-NGC6412, 
the luminosities of the plateau are the same. 
The length of SN-NGC6412 plateau is uncertain due to the lack of observations. %, but it was probably similar to that of SN~2005cs. 

The light curves are also compared with those of other Type II-P SNe collected from the literature. 
One of them is SN~1997D (\citealt{Turatto-ecapture}, \citealt{benetti-1997d}) which is the prototype of low luminosity subclass. 
Unfortunately the light curve of SN~1997D is very sparsely sampled.  
SN~2009md is also a subluminous SN, which had a low mass progenitor similar to SN~2005cs (\citealt{Fraser-2009md}).

SN~1999em (\citealt{leonard-1999em}), SN~2004et (\citealt{sahu-2004et}) and SN~2012aw (\citealt{bose-2012aw}) on the other hand, 
are well-observed, normal II-P supernovae. They serve as a control group for the modeling program used below (Section 8).

\begin{table*}
\begin{center}
\begin{tabular}{ l c c c c c c c }
\hline

{\bf SN }          & {\bf Explosion }       & {\bf Distance }        & {\bf E(B-V) }       & {\bf Reference } \\
                   & {\bf epoch [MJD] }                 & {\bf [Mpc] }           & {\bf [mag] }        & {\bf  } \\
\hline
{\bf 1997D }       & 50361.0*                     & 13.4                   & 0                   & \citealt{benetti-1997d}; \citealt{spiro-sublum}\\
{\bf 1999em }      & 51480.4                      & 11.7                   & 0.1                 & \citealt{leonard-1999em}; \citealt{leonard-1999em-dist}\\ 
{\bf 2004et }      & 53270.0                      & 5.6                    & 0.41                & \citealt{sahu-2004et}\\ 
{\bf 2005cs }      & 53549.0                      & 7.1                    & 0.05                & \citealt{pastorello-2005cs}; \citealt{Takats-05csref}\\ 
{\bf 2009md }      & 55162.0                      & 21.28                  & 0.1                 & \citealt{Fraser-2009md}\\
{\bf 2012aw }      & 56002.1                      & 9.9                    & 0.074               & \citealt{bose-2012aw}\\
{\bf SN- }        &                              &                        &                     &  \\
{\bf NGC6412 }     & 57210.0                      & 17.85                  & 0.035               & This work \\%\citealt{ref-NED}\\

\hline
\end{tabular}
\caption{Adopted parameters. * SN 1997D explosion epoch is uncertain. 
%\citealt{benetti-1997d} gives 50430 $\pm$ 20 for the maximum. 
We adopted the value given by (\citealt{spiro-sublum}) with extended plateau range.
Explosion epoch discuss is can be found in Section 6, distance discuss in Section 4, and extinction estimation is in Section 3. }
\label{Adopted}
\end{center}
\end{table*}

%1997D E(B-V) \citealt{sahu-2004et}-ban van
%1999em distance is innen lett megtalalva, az elter az eredeti cikktol, egy kesobbi cikk teszi helyre
%SN 2005cs cikkben 0.05 az extinkcio, de 0.03 volt hasznalva (ned), nem nagy kulonbseg
%SN 2005cs tavolsag Takas es Vinko cikkbol van

\section{Spectroscopy}

\begin{figure*}
\begin{center}
\includegraphics[width=\columnwidth]{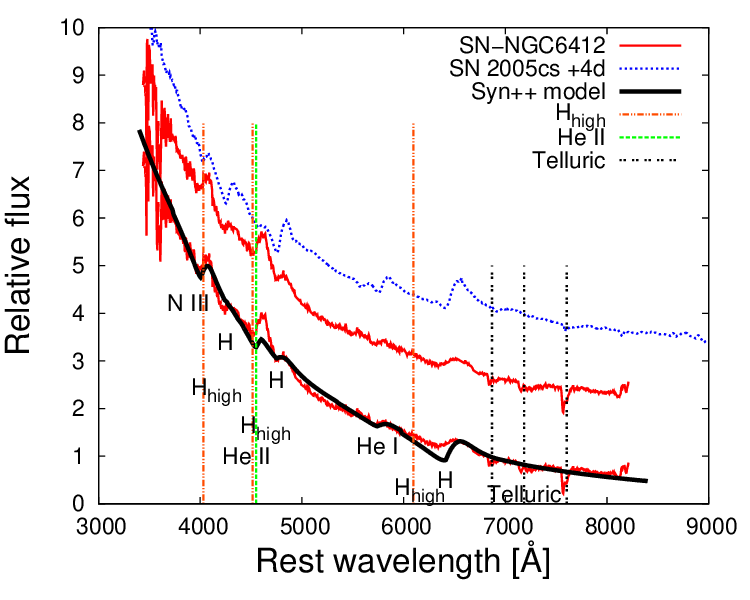} 
\includegraphics[width=\columnwidth]{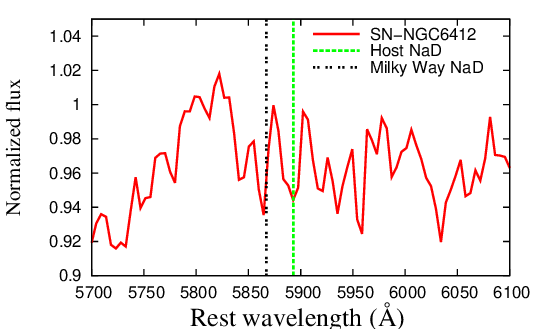} 
\includegraphics[width=\columnwidth]{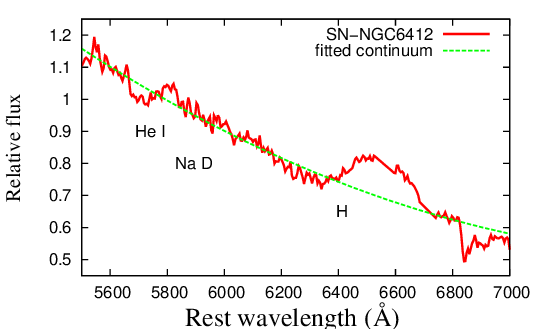} 
\includegraphics[width=\columnwidth]{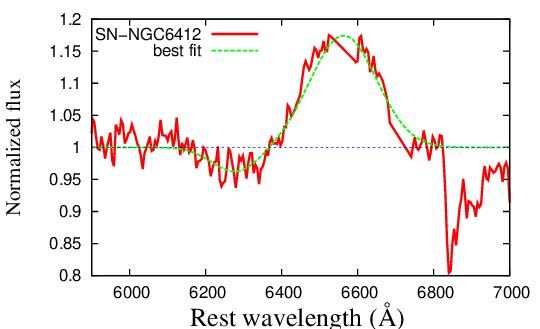} 
\caption{Upper left: The spectrum of SN-NGC6412 corrected for redshift, the best fitting \texttt{Syn++} model, and the spectrum of SN 2005cs 
(shifted vertically for easy visual comparison)
The feature at 4578 \AA\;can be fitted with high velocity H (23\,000 km s$^{-1}$) and with H II. See text.
%Bottom: The spectrum of SN-NGC6412 corrected for redshift, the best fitting \texttt{Syn++} model, and the spectrum of SN 2005cs 
%(shifted vertically by and arbitrary amount for easy visual comparison).
Upper right: The Na D line in SN-NGC6412 spectrum.
Bottom left: The region around the $H_{\alpha}$ in SN-NGC6412 spectrum.
Bottom right: The P Cygni profile of the $H_{\alpha}$, and the fit used to calculate the equivalent width, see text.
}
\label{spectrum}
\label{NaD}
\end{center}
\end{figure*}

There is only a single observed spectrum available for SN-NGC6412 in the literature (\citealt{toma-spectra}) 
taken one day after the discovery (57214 MJD). It is shown in Fig.~\ref{spectrum} top left panel.
The spectrum is contaminated by telluric lines as well as emission features from the host galaxy.
Before further analysis, the following host galaxy features were removed from the spectrum: 
\ion{H}{I} $\lambda\lambda$ 4341, 4861, 6563, \ion{N}{II} $\lambda\lambda$ 6548, 6583, \ion{O}{I} $\lambda\lambda$ 3727,
\ion{O}{III} $\lambda\lambda$ 4959, 5007, \ion{Mg}{II} $\lambda\lambda$ 2798, \ion{S}{II} $\lambda\lambda$ 6717, 6731, Ca H\&K.
%the telluric lines were ignored during the fitting.

We utilized the \texttt{Syn++} modeling code (\citealt{thomas-syn}) to fit the observed spectrum after correcting it for extinction and redshift.
\texttt{Syn++} is an advanced version of \texttt{SYNOW} which uses the Sobolev approximation (see \citealt{fisher-syn}) 
to calculate the P Cygni features formed by pure resonance scattering in a homologously expanding SN atmosphere.

The spectrum of SN-NGC6412 (taken on 2015-7-11, 57214 MJD) shows H, \ion{He}{II}, \ion{He}{I} and \ion{N}{III} lines, 
which elements we expect in an early hydrogen-rich supernova atmosphere (see \citealt{hatano-syn}, Fig. 2a). 
The parameters used for the fitting are shown in Table~\ref{specfit}. %Fe~I
The feature at 4578 \AA, can be either He~II or high velocity H (23\,000~km s$^{-1}$). 
However a high velocity H would imply a remarkable absorption around 6095 \AA, 
which is not the case (Fig.~\ref{spectrum}). That is why we adopted He~II.
This is also consistent with the temperature. The formation of the He~II line requires high temperature.
The temperature is 19\,000 $\pm$ 3000~K, while
the photospheric velocity of the best-fit model is 7000~km s$^{-1}$,  
which implies a very early stage.
The spectrum comparison made with the SNID (Supernova Identification, \citealt{tonry-snid})
reveals the spectrum of SN~2005cs, 
taken on 2005-07-01, 2005 (4~days after the shock breakout), being the most similar one. 
However, SN 2005cs lacks the strong $\lambda 4578$ feature that is apparent is SN-NGC6412 (attributed to \ion{He}{II}, see above).
At that epoch the photospheric velocity of SN~2005cs was 6950~km s$^{-1}$, while its temperature was 13350~K (\citealt{dessart-2005cs})
As we can see the expansion velocity is very similar to the SN-NGC6412. 

The difficulties of determining the continuum of a noisy spectrum render any measurement of equivalent widths very uncertain.
Nevertheless, we performed such a procedure for the P Cygni profile of the H$\alpha$ line, 
in order to have some rough quantities for comparison with other supernovae (Fig.~\ref{spectrum}, bottom two panels).
The equivalent widths are $5 \AA$ for the absorption part and $37.2 \AA$ for the emission part,  %full: 39.97 +- 1.4 A
giving an absorption/emission ratio $a/e = 0.13$.
For comparison, SN2005cs has the corresponding values of 
$9.8 \pm 1 \AA$ and $79.5 \pm 1.1 \AA$  for absorption and emission respectively, and a ratio $a/e = 0.12$.  %full: 89.317 +- 1.5 A
The $a/e$ is similar for the two, however for SN-NGC6412 it may be higher due to the uncertainty of the continuum.
The SNe start with low $a/e$ which then increase over time as the temperature drops (see \citealt{gutirez-apere}).
SN-NGC6412 has a smaller equivalent widths than the SN~2005cs. This is because of the higher temperature.

The upper limit of the velocities for SN-NGC6412 reaches $30\;000$ km s$^{-1}$ (see Table~\ref{specfit}), 
resulting in a longer shallow absorption. %nagyobb hullamhosszu
However \texttt{Syn++} fit shows that this feature is the part of the outer atmosphere, and not a high velocity Hydrogen feature. 
%Fig.~\ref{Vhas} shows the photosphere velocities of the SNe collected from the literature 
%(1999em: \citealt{leonard-1999em}; 2005cs: \citealt{dessart-2005cs}; 2009md: \citealt{Fraser-2009md}; 2012aw: \citealt{bose-2012aw}).

%\begin{figure}
%\begin{center}
%\includegraphics[width=\columnwidth]{eps/PSN-NGC6412_V_HAS.eps} 
%\caption{Photospheric velocities of the SNe.
%Filled symbols represent the results from the SED fitting (see text), while open symbols correspond to values collected from literature.
%(1999em: Leonard, 2012aw: Bose, 2005cs: Dessart, 2009md: Fraser, 
%(SN-NGC6412 value is from the \texttt{Syn++} fitting of the spectrum.)}
%\label{Vhas}
%\end{center}
%\end{figure}

\begin{table*}
\begin{center}
\begin{tabular}{ l c c c c c }
\hline

Parameter                   & Photosphere & H     & He II & He I & N III  \\
\hline
Velocity [km s$^{-1}$]      & 7000        & 7000  & 9000  & 7000  & 7000  \\
MAX Velocity [km s$^{-1}$]  & 30000       & 30000 & 30000 & 30000 & 30000 \\
Temperature [K]             & 19000       & 19000 & 19000 & 19000 & 19000 \\
$\log(\tau)$                & -           & -0.4  & -1.2  & -1.5  & -1.3  \\
aux [km s$^{-1}$]           & -           & 8000  & 8000  & 8000  & 8000  \\

\hline
\end{tabular}
\caption{Parameters used to fit SN-NGC6412 spectrum with Syn++. }
\label{specfit}
\end{center}
\end{table*}

%km/s

\section{The evolution of temperature and radius at the photosphere}

Using the photometry of SN-NGC6412, we determined the temperature ($T$) and radius ($R$) of the photosphere of the SN.  
The spectral energy distribution (SED) is modeled as a diluted blackbody radiation, using the following formula:
\begin{equation}
F_{\lambda} = (R/d)^2 \cdotp \xi^2 (T) \cdotp \pi \cdotp \mbox{B}(\lambda,T) \cdotp 10^{-0.4 \cdotp A_{\lambda}}
\label{dilBB}
\end{equation}
where $F_{\lambda}$ is the flux, $R$ is the radius, $d$ is the distance, $\lambda$ is the wavelength, B is the Planck function, $A$ is the extinction, 
and $\xi (T) $ is the dilution factor (\citealt{eastman-diluted}; \citealt{hamuy-diluted}; \citealt{dess-hill-BB}) 
that corrects the flux of a pure blackbody to the one formed in a strongly scattering SN atmosphere. 
$\theta = R / d$ is the angular radius of the photosphere.

\begin{figure*}
%\begin{center}
\includegraphics[width=\columnwidth]{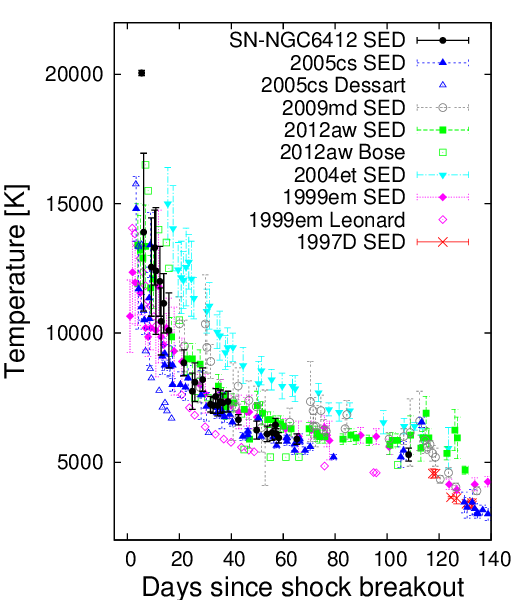} 
\includegraphics[width=\columnwidth]{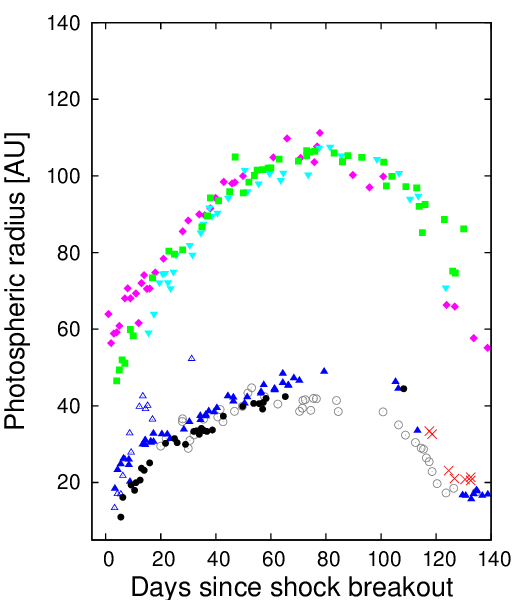} 
\caption{Evolution of the photospheric temperatures (left) and radii (right) of the discussed SNe.
Filled symbols represents the results from the SED fitting (see text), while open symbols corresponds to values collected from literature.
The first point of SN-NGC6412 gives $\sim$ 20\,000~K which is consistent with the value comes from the spectrum fit (19\,000~K)}
\label{Thas}
\label{Rhas}
%\end{center}
\end{figure*}

The fitting of Equation (\ref{dilBB}) was also done to all of the other SNe used for comparison.  
Fig.~\ref{Thas} show the $T$ and $R$ values obtained in this way. %of the SNe. %and \ref{Rhas}
The B band flux was not included in the blackbody fitting, because the blackbody flux in the blue and UV region significantly differs
from the observed fluxes.
However the photospheric temperature and radius values for the reference SNe in the literature were computed including
the B flux in the fitting.
This explains the slightly different values obtained for the same SNe, as seen in Fig.~\ref{Thas}.
Nevertheless all of the SNe seem to evolve in a similar way: the initial high temperature
decreases very quickly, and at the end of the plateau phase it becomes quasi-constant around 7000 K,
which is close to the recombination temperature of the H.
The radius shows a linear expansion at the beginning, but after a while it also becomes quasi-constant,
and dropping down at the end of the plateau.
The temperature of SN-NGC6412 evolves similar to the other SNe, except for the first point which gives 20\,000 K.
This suggest a very early epoch.
SN-NGC6412 and SN~2005cs have significantly lower photospheric radii, suggesting low photospheric velocities.
Also, this might be the reason why they are very dim.\\

The time of shock breakout ($t_0$) is an important required parameter for the modeling of the LC.
%The SN-NGC6412 was discovery on 57213~MJD, so 
Naturally $t_0$ must be earlier than the discovery epoch (57213 MJD).
Epochs significantly earlier than the discovery are unlikely because the temperature on the first observed epoch was quite high (20\,000 K see Fig.~\ref{Thas}),
implying that it is a very early epoch, even down to only one day. The mean of these two value is 57210.5.
Also, if $t_0=$ 5210 MJD, the photosphere radius and the bolometric LC of SN-NGC6412 and SN~2005cs become very similar (see Section 7).
Therefore we adopt $t_0 = $ 57210$ \pm 2$~MJD as the epoch for the shock breakout.
Fortunately the uncertainty is quite low, making it adequate for the modeling.

\section{Bolometric light curve}

The modeling code requires the bolometric light curve as an input.
Unfortunately, there are only optical data available. To handle this we have to make estimations for the regions not covered by the observations. \\
The total flux in the infrared (IR) region, covering wavelengths beyond the I~band, was calculated as the integral of the fitted diluted black body radiation.
The diluted black body is only fitted to bands that are red-ward from the B-band,
because the blue part of the spectrum significantly differs from that of black body radiation.

\begin{figure}
\begin{center}
\includegraphics[width=\columnwidth]{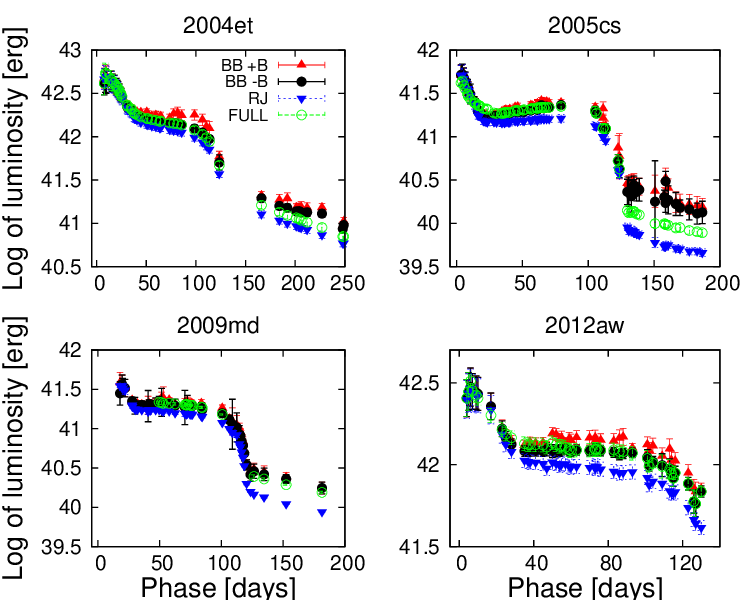} 
\caption{Comparing the methods to calculate bolometric flux. 
The Figure only show those SNe which has both UV and IR data available beside the optical data.
BB: the fitted blackbody, which was computed with (+B) and without (-B) including the B-band flux. RJ: Rayleigh-Jeans approximation.
FULL: Direct integral using JHK and Swift (where available) fluxes. Optical regime calculated with direct integral in any case.
UV regime calculated with interpolation (see text) where Swift flux was unavailable.}
\label{BBhas}
\end{center}
\end{figure}

We tested the effect of the inclusion/omission of the B-band flux into/from the blackbody fitting in the following way. 
For those SNe that had $JHK$ data available, 
the flux integral in the near-IR regime was also calculated via direct integration of the $JHK$ fluxes serving as a control. 
In addition, the IR flux was also estimated by fitting a Rayleigh-Jeans (RJ) tail to the I-band flux and integrating it 
between the I-band and infinity.
These results (the direct integration and the RJ approximation) were 
then compared to the integral of the fitted blackbody, 
which was computed with and without the B-band flux. 
It turned out that omitting the B-band flux from the blackbody fitting (i.e. fitting  only to the $VRI$ bands) 
gives much closer values to the result from the direct integration, far better than the RJ approximation, 
at least during the photospheric phase.
Fig.~\ref{BBhas} shows the results of the testing.  
The green-coloured open symbols depict the actual integration using UV, optical and IR data, serving as a control.
The plot only includes points with either UV or IR data available beside the optical data
%This only shows those points where are at least UV or IR data was available beside optical 
(SN 2012aw only had optical at the tail phase, thats is why it is missing from the plot).
The filled black circles show the estimation excluding the B band, the red filled triangles show the estimation with B band,
and the blue filled upside down triangles show the RJ approximation.

This method does not work in the nebular phase, 
because the blackbody assumption breaks down after the plateau as the ejecta becomes more and more transparent. 
Thus, after the plateau phase we applied only the direct integration of the near-IR fluxes, except for SN~1999em, 
where we added the integral of a Rayleigh-Jeans tail fitted to the observed I-band fluxes. 
The same technique was applied to the single epoch $BVRI$ fluxes of SN-NGC6412 taken after the end of the plateau phase.  

The flux in the ultraviolet (UV) region, encompassing wavelengths shorter than band B, was extrapolated to $2000$~\AA\/ from the B and V bands, 
and assumed zero for even shorter wavelengths (\citealt{lyman-bc}).
For the optical region we integrated the observed and de-reddened fluxes over the spectral bands using the trapezoidal rule.
Adopted distance and extinction values are shown in Table~\ref{Adopted}.
\\

The data for the comparison SNe were collected from the following sources. 
SN~2005cs: optical and NIR: \citet{pastorello-2005cs}; UV: \citet{Brown-05csrefUV}. 
SN~2012aw: optical: \citet{bose-2012aw}; \citet{Dallora-12awref}; 
NIR (dates preceding the nebular phase): \citet{Dallora-12awref}; UV: \citet{Bayless-12awUV}. %main opt: \citealt{bose-2012aw} + add opt: \citealt{Dallora-12awref}
SN~2004et: optical: \citet{sahu-2004et}; NIR (after the tail): \citet{Maguire-04etrefNIR}. %another data: \citealt{Misra-04etref}, and there's another Sahu
SN~1999em only has existing optical data: \citet{leonard-1999em}. 
%this data used, more data, and bolometric: \citealt{Elmhamdi-99emref}
%UV data: SWIFT UVOT, NIR: near-infrared: J H K bands.
SN~2009md: optical and NIR (after the tail): \citet{Fraser-2009md}.\\

\begin{figure*}
\begin{center}
\includegraphics[width=\columnwidth]{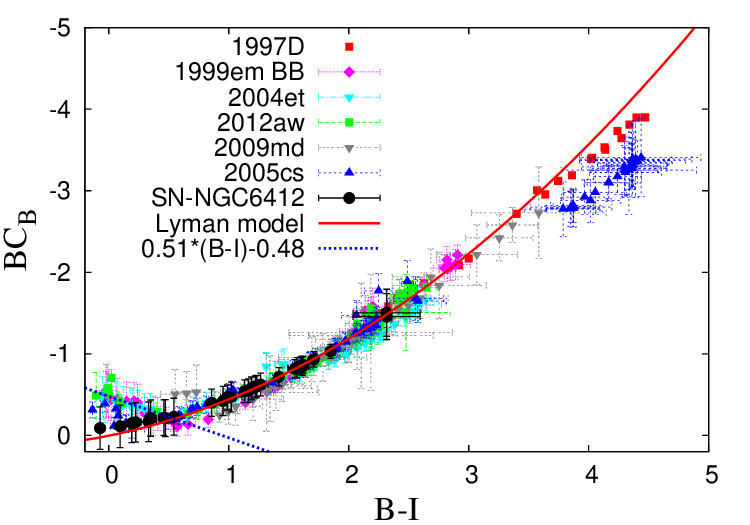} 
\includegraphics[width=\columnwidth]{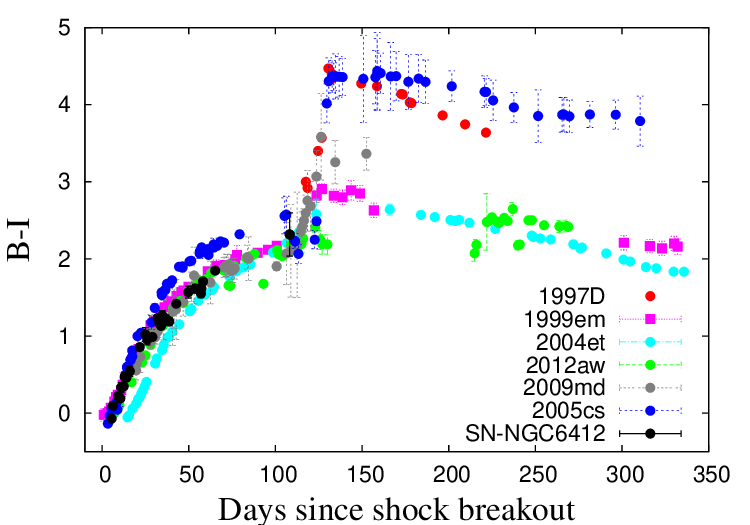} 
\caption{Left: The $\text{BC}_B$--$(B-I)$ relation for the SNe in this study:
%the resultant line from the two model (Lyman model and $0.51 \cdotp (B-I) - 0.48$) 
%using always the smaller $\text{BC}_B$ value (bigger correction).
For $B-I$<0.8 we apply the prediction from the linear fitting (blue dotted line) instead of the fitted polynomial (red curve). 
Right: the colour (B-I) development over time.
}
\label{Szinhas}
\end{center}
\end{figure*}

The errors for the bolometric fluxes were calculated as follows.
The optical and UV flux errors were estimated from the error propagation while integrating the flux with the trapezoidal rule. 
The IR flux error comes from the uncertainty of the temperature of the fitted blackbody. 
However, this does not include the uncertainty due to the validity of the applied blackbody model. 
Thus, in order to include this kind of uncertainty in the total error budget, 
we added a fixed percent of the UV and IR flux to the formal errors mentioned above.  
For the IR the percentage depends on the goodness of fit of the diluted blackbody: we define it as the flux
ratio of the direct integral of the optical fluxes 
and the integral of the blackbody in the optical region.
For UV it was fixed as 50\%. With this value, the uncertainty of the bolometric magnitude is 0.3 mag at the beginning of the LC.
This is the approximate error in the $\text{BC}_B$--$(B-I)$ relation around $B-I = 0$ (see Fig.~\ref{Szinhas}).

Gaps in the optical band were filled in by linear interpolation. 
Bolometric fluxes were generated only for measurements involving at least 3 optical bands.\\

To check the correctness of this method, we used an empirical correlation found by \citet{lyman-bc} between the bolometric correction of 
the $B$ band ($\text{BC}_B =$ bolometric magnitude minus B-band magnitude) and the color index $B-I$. 
They modeled the correlation by fitting a second order polynomial to it.
We plotted $\text{BC}_B =$ vs $B-I$ for every SN in Fig.~\ref{Szinhas}. 
The plot shows that SN~2005cs and the other SNe closely follow this model, and SN-NGC6412 also closely follows this if
the extrapolation is limited to $\lambda \geq 2800$~\AA , instead of $2000$~\AA. 
This means that the two methods (integral, and the Lyman model) give the same result.

\begin{figure}
\begin{center}
\includegraphics[width=\columnwidth]{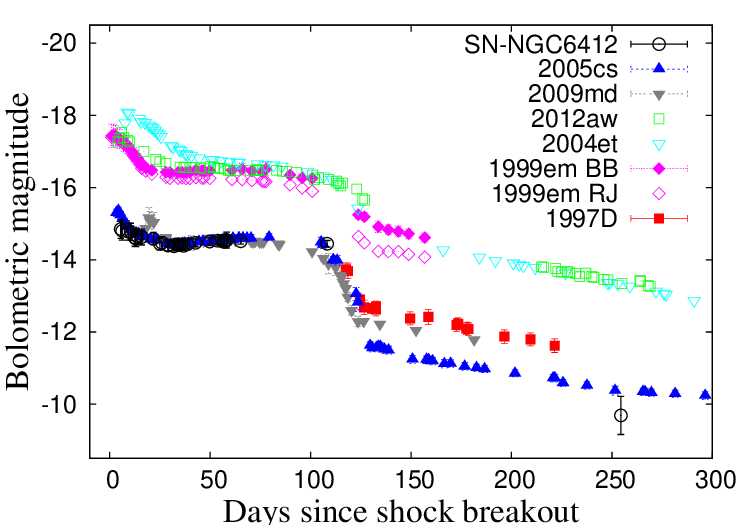} 
\caption{Bolometric light curves of the SNe. 
%\textsl{Full symbols}\/: integrated, \textsl{hollow symbols}\/: 
%calculated from the $\text{BC}_B$--$(B-I)$ relation (Fig.~\ref{Szinhas}). Note: after about 20 day, the two method fully overlaps.
} 
\label{Bolhas}
\end{center}
\end{figure}

However the very early UV flux is still highly unreliable and the Lyman model does not fit it well. 
SN~2005cs and SN~2012aw have existing UV observations and they were used to test the UV region. 
We fitted the deviation from the Lyman model with a straight line:
\begin{equation}
\text{BC}_B = 0.51 \cdotp (B-I) - 0.48 \qquad\text{if }(B-I) < 0.54
\end{equation}
The other SNe with the extrapolated UV flux show the same deviation from the Lyman model at the beginning, except for SN-NGC6412. 
The bolometric light curves corrected in this way are shown in Fig.~\ref{Bolhas}. 
In the literature, RJ approximation was used for SN~1999em instead of blackbody fitting (\citealt{Elmhamdi-99emref}).
This approximation is adequate for the nebular phase, but not for the photometric phase, as can be seen in Fig.~\ref{BBhas}.
Thus, we used the better BB fitting method (see Fig.~\ref{Szinhas}, and Fig.~\ref{Bolhas}).
\\

Note that in Fig.~\ref{Szinhas} SN 2005cs, and to some extent SN 1997D and 2009md deviate from the Lyman model in the high $B-I$ values.
The same figure also shows the development in time of the $B-I$ colour. At first all SNe start with $B-I \sim 0$, 
meaning that the ejecta is blue in colour. As the temperature drops over time (Fig.~\ref{Thas}), the $B-I$ colour increases.
At the end of the plateau it suddenly rises. The LL SNe have an even higher rise of $B-I$ than regular Type II-P SNe,
reaching $B-I = 4.5$ mag. This is a unique feature of LL SNe.
The sample made by \citet{lyman-bc} was only fitted till about $B-I =  \sim 3$ mag. This model at $B-I = 4.5$ mag is not entirely valid, however, it is 
more or less still adequate.
%Lymanban irjak is, hogy: it is valid over the B-I colour range -0.4 -- 2.3 (0.0 -- 2.8).

\section{Bolometric light curve fitting, and the modeling code}

We used an upgraded version of the LC2.2 semi-analytic light curve code (see \citealt{Nagy-Vinko-LC}) to fit the quasi-bolometric light curve. 
This code is based on a model that was originally described by \citet{Arnett-Fu-LC} and later extended by 
\citet{Popov-LC}; \citet{Blinn-Popov-LC} and \citet{Nagy-LC}. % to model the light curves of CCSNe.
A wide variety of SN light curves can be modeled with this code, depending on the choice of the initial parameters, 
such as the ejected mass ($M_\text{ej}$), the initial radius of the progenitor ($R_0$), the total explosion energy ($E_0$), 
and the mass of the synthesized $^{56}$Ni ($M_\text{Ni}$) which at later phases directly determines the emitted flux.
The model assumes a homologously expanding and spherically symmetric SN ejecta having a uniform density core and 
an exponentially decreasing density profile in the outer layers. 
The diffusion approximation was used for the radiation transport. The recombination causing the rapid change of the effective opacity 
in the envelope is taken into account in a very simple form (see \citealt{Arnett-Fu-LC}, Eq.~(A8)). 
%($\kappa$ if $T>T_{ion}$, zero otherwise)

The LC2.2 code is only a modeling tool, thus it does not contain any routine for numerically fitting the output model to the observed light curves. 
Thus we added a 
Markov Chain Monte Carlo (MCMC) method 
using the Metropolis-Hastings Algorithm with Gibbs sampler 
(\citealt{Metropolis-mcmc}; \citealt{Hastings-mcmc}; \citealt{Gilks-mcmc}), to find the best fits to the bolometric light curve
(which are selected by the least squares method, $\chi^2 = \Sigma_i^N (\frac{M_i-D_i}{\sigma_i})^{2}$, where M: model, D: data, $\sigma$: measurement errors), 
explore the parameter space, and investigate correlations between the parameters.

The MCMC program samples the joint posterior probability distribution of the parameters, 
given suitable \textsl{a priori} probability distributions for each parameter and the goodness of fit, 
$\chi^2$, which establishes the \textsl{likelihood} as ${\cal L} = \exp(-\chi^2/2)$ (We assume Gaussian errors as usual).

The accepted maximum and minimum parameter regions are based on \citet{hamuy-parameters}, however the range has been extended to be more general.

Because of the direct sampling of the parameter space, any correlation between the parameters can also be readily observed.
There are two known main parameter correlations (\citealt{Arnett-Fu-LC}, \citealt{Nagy-LC}): 
between $M_\text{ej}$ and $E_\text{kin}$, and between $R_0$ and $E_\text{th}$.  
The opacity ($\kappa$), and the exponent of the power-law density profile (of mass) ($s$) may also be sampled, 
which are also known to have correlations (\citealt{Nagy-Vinko-LC})).
There are other parameters not entering into the sampling process, i.e. they are not fitted. 
These include the ionization temperature ($T_\text{ion}$), the date of explosion ($t_0$), and the distance ($d$), 
$t_0$ and $d$ can be determined independently, and $T_\text{ion}=5500K$ was adopted as the ionization temperature of the hydrogen.
(Note: if $t_0$ and $d$ are not well known, the algorithm, after including them into the fitting process, 
is also able to determine their uncertainties.)
Using constant density model $s=0$ gives a good agreement with the hydrodynamical models, so $s=0$ was adopted.
The opacity $\kappa$ also correlates with $E_\text{kin}$ and $M_\text{ej}$. %(see Section 10). 
%(More about this latter in section 'Correlations'.)
%
Because of this we have also sampled $\kappa$ between
$\kappa=0.3$ cm$^2$ g$^{-1}$ and $\kappa=0.2$ cm$^2$ g$^{-1}$ (with uniform \textsl{prior}: $\kappa=[0.2:0.3]$ cm$^2$ g$^{-1}$), 
These are the average opacities calculated by the public code SNEC (\citealt{Morozova-SNEC}; see \citealt{Nagy-Vinko-LC} for details).

Because the light curve sampling is not very dense, to say the least, we had to take some measures to ensure a correct fit of the model on the fluxes.
The weights of the last two points have been increased (their uncertainties decreased 10 times), 
%The increase of the weights was required 
to ensure that the these important points are properly fitted 
(in other words, the model goes through or close to these important points).

Also, the first 30~days after the explosion are not fitted, because the bolometric fluxes are 
uncertain in that regime, due to a lack of UV flux measurements.
Luckily this phase of the light curve is caused by the ejected outer shell and not the core, 
and this shell is independent from the core as it has different physical parameters, for example larger radius, so it must be fitted separately.
(see \citealt{Nagy-Vinko-LC} for more information).

The error of $t_0$ is very low so it does not affect the other parameters significantly.
The tail of SN-NGC6412 seems to be 4 magnitude dimmer than at the end of the plateau, 
just like in the case of SN~2005cs. This means $\sim$35 times lower flux, 
so it adds a very small portion of energy to the luminosity.

\section{The missing plateau endpoint and its implications}

\begin{figure*}
\begin{center}

\includegraphics[width=1.1\columnwidth]{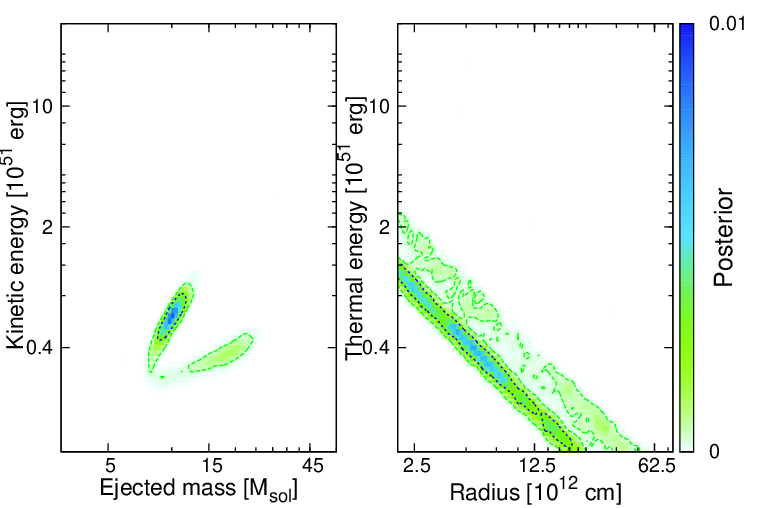} 
\includegraphics[width=0.9\columnwidth]{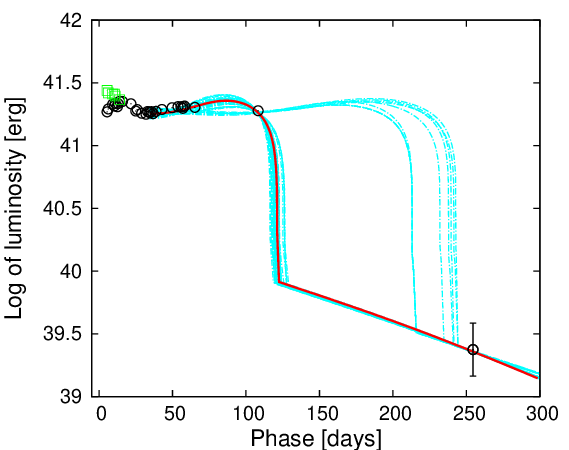} 
\includegraphics[width=1.1\columnwidth]{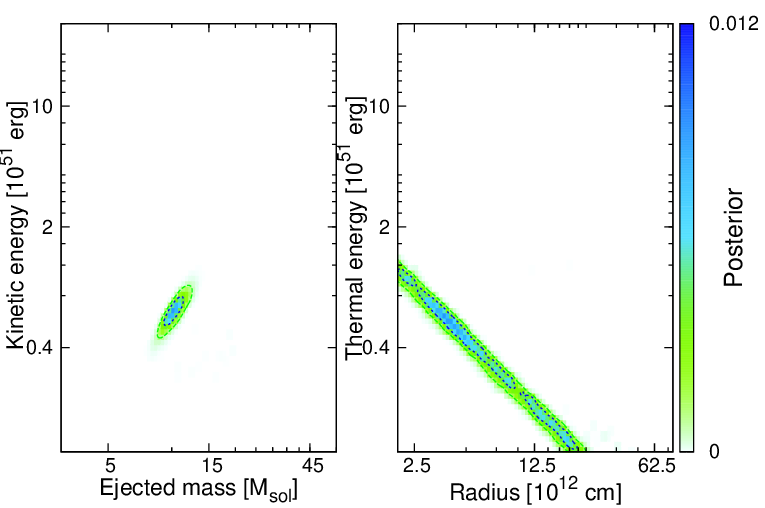} 
\includegraphics[width=0.9\columnwidth]{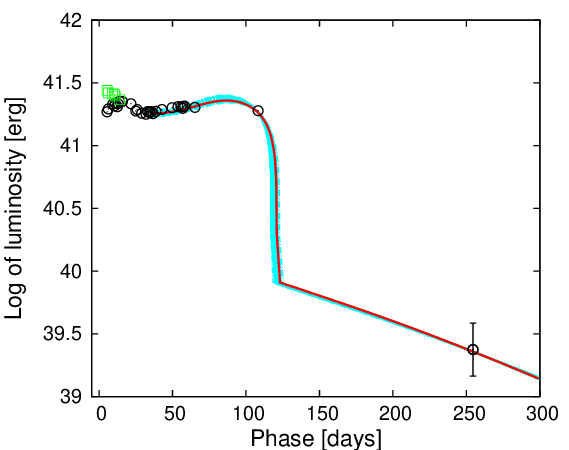} 
\caption{Upper left: probability distributions of SN-NGC6412 joint parameters, $d=17.85 \;\text{Mpc}$, $s=0$, $\kappa=[0.2:0.3]$ cm$^2$ g$^{-1}$. 
%The colors encode the likelihood of the individual fits.
The contours show the confidence interval of 67\% and 95\%. Note that there is a second (multi-modal) pole.
Upper right: SN-NGC6412 accepted fits (within 95\% probability contour) including the second (multi-modal) pole. The red line is the best fit. $d=17.85 \;\text{Mpc}$, $s=0$, $\kappa=[0.2:0.3]$ cm$^2$ g$^{-1}$. 
The green symbols shows the added UV from the $BC_B(B-I)$ relation (Fig.~\ref{Szinhas}).  The second (multi-modal) pole fits have the end of the plateau too far, over +200 day, which is unrealistic. 
Bottom left: The probability distributions with given velocity prior. The second (multi-modal) pole vanished.
Bottom right: SN-NGC6412 accepted fits with added velocity priori. Note that the bad fits vanished.
} 

\label{PSN_space}
\label{fitrossz}
\label{PSNspaceJO}
\label{PSN_best}
\end{center}
\end{figure*}

The code was applied to the data of SN-NGC6412 and to the other 
comparison/reference SNe. The posterior distributions and the best fits of SN-NGC6412 
are shown in (Fig.~\ref{PSN_space}) ($d=17.85$~Mpc, $\kappa=[0.2:0.3]$ cm$^2$ g$^{-1}$).
The posterior distributions of the reference objects SN~1999em, SN~2004et, SN~2005cs and SN~2012aw
are shown in the Appendix. 
The detailed discussion of the results is given in Section 10.

The unfortunate lack of an observed plateau endpoint is a serious issue. 
The light-curves of SN 2005cs and SN-NGC6412 are running closely together up to the last plateau observation (Fig.~\ref{Bolhas}). 
Luckily the MCMC algorithm samples the whole parameter space and searches for all possible solutions with different plateau endpoints.
The shape of the observed data constrains the endpoint position.
In our case the absence of endpoint observation gives rise to a secondary maximum (a case of multi-modal distribution)
in the parameter space, a solution with larger $M_\text{ej}$.
A comparison of the model light-curves of the two maxima, which can be seen on (Fig.~\ref{fitrossz}), 
clearly shows that the solution corresponding to the second maximum is far worse.
The endpoint of the plateau is also too far away.
A comparison with other LL~SNe also reveals that such a late endpoint is very unlikely, 
see also Table~\ref{LLSNe} showing the plateau end epoch and luminosity at the 50th day.
The plateau end epoch is between day 100 and 130, while the luminosity difference is almost a factor of 10.

The spectrum of SN-NGC6412 is also similar to that of SN 2005cs at the similar epoch, having nearly the same early photospheric 
velocity, which suggests that they have similar expansion velocities at their outmost layers. %(Fig.~\ref{Vhas}).
The development of the photospheric radius (Fig.~\ref{Rhas}, right panel) is also very similar.
%Consequently the development of the photospheric velocity cannot be radically different. 
So we repeated the fitting while including also the expansion velocity with a Gaussian prior. 
We assume a homologously expanding spherical symmetric ejecta.
The expansion velocity is a calculated parameter, 
which stands for the velocity of the outermost layer of the SN ejecta (see \citealt{Nagy-Vinko-LC}), 
and assumed to be constant.
As the photosphere propagates inward the remnant, it will be located in a different radius resulting different photosperic velocities.
The expansion velocity for SN-NGC6412 was determined from the SN 2005cs fit
but with larger uncertainty: $3500$ km s$^{-1}$ mean and $2\sigma = 500$ km s$^{-1}$ 
(see Appendix %Table~\ref{2005cs_fitvalue} 
for the exact values for SN 2005cs).
After repeating the fitting with these modifications, the second maximum vanished and only the better one remained, as shown in Fig.~\ref{PSNspaceJO}. 
Note even with this, the uncertainty of the plateau end is still rather high ($\sim$20 day), 
however this uncertainty is included in the fit and the final results.

\begin{table*}
\begin{center}
\begin{tabular}{ l c c c c c }
\hline
{\bf Name} & {\bf log L [erg s$^{-1}$] \@@ day 50 } & {\bf Plateau end [d] } & {\bf $M_\text{Ni}$ [$0.001 M_{\odot}$]  } & {\bf Ref.}\\
\hline
SN 1997D	& -	& 125		& 5 $\pm$ 4	& 2, 3, 4 \\
SN 1999br	& 40.60	& -		& 2 $\pm$ 1	& 1 \\
SN 1999eu	& -	& 100-120	& 1 $\pm$ 1	& 1, 4 \\
SN 2001dc	& 40.85	& 110		& 5 $\pm$ 2	& 1 \\
SN 2003Z	& 40.90	& 120		& 5 $\pm$ 3	& 4 \\
SN 2005cs	& 41.10	& 120		& 3 $\pm$ 1	& 5 \\
SN 2008bk	& 41.25	& 130		& 7 $\pm$ 1	& 6, 7 \\
SN 2008in	& 41.35	& 105		& 12 $\pm$ 5	& 8 \\
SN 2009N	& 41.50	& 110		& 20 $\pm$ 4	& 9 \\
SN 2009md	& 41.05	& 115		& 4 $\pm$ 1	& 10 \\
SN 2010id	& 40.75	& 120		& -		& 11 \\
SN-NGC6412	& 41.10	& 110-130	& 1.5 $\pm$ 0.8 & this work\\
\hline
\end{tabular}
\caption{ The LL SNe family features. Ref.: (1) \citet{Pastorello-sublum}; (2) \citet{Turatto-ecapture}; (3) \citet{benetti-1997d};
(4) \citet{spiro-sublum}; (5) \citet{pastorello-2005cs}; (6) \citet{Mattila2008}; (7) \citet{Vandyk-12awrefHST};
(8) \citet{Roy2011}; (9) \citet{Takats2014}; (10) \citet{Fraser-2009md}; (11) \citet{Gal-Yam2011};
}
\label{LLSNe}
\end{center}
\end{table*}
%(13) \citealt{Maguire-04etrefNIR}

\section{Results}

Fig.~\ref{PSNspaceJO} shows that our results are consistent with
\citet{Arnett-Fu-LC} and \citet{Nagy-LC}, regarding the correlations between the parameters.

Due to the correlation between $R_0$ and $E_\text{th}$, they cannot determined separately, only their product,
$R_0 \cdot E_\text{th}$, can be derived, and should be used as an independent parameter 
in subsequent analyses, instead of $R_0$ and $E_\text{th}$.

\begin{figure}
\begin{center}
\includegraphics[width=\columnwidth]{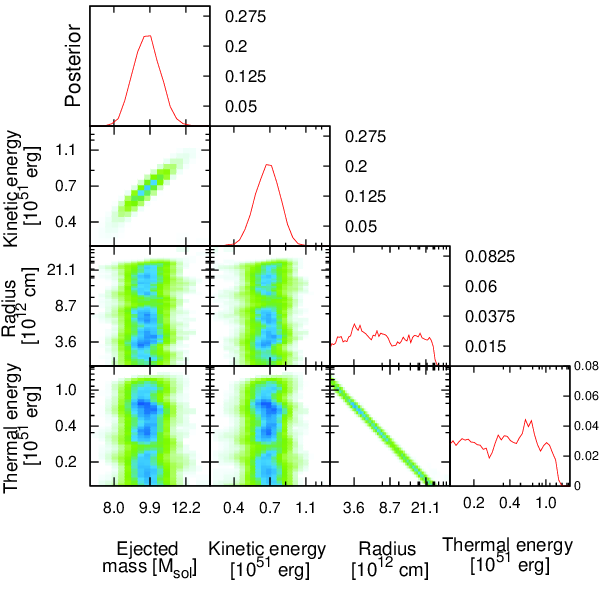} 
\caption{Correlations between the main parameters shown of SN-NGC6412 with added velocity prior. 
The color shows the probability distributions of the joint parameters.
Here $s=0$, $\kappa=0.3$ cm$^2$ g$^{-1}$ were chosen.} 
\label{spaceALL}
\end{center}
\end{figure}

For $M_\text{ej}$ and $E_\text{kin}$, their correlation
shows a parabolic trend (in logarithm).
This is a more complex correlation than we previously thought.
We can not make an independent parameter from $M_\text{ej}$ and $E_\text{kin}$ because of this complex correlation trend.
They are however, less significantly correlated (Fig.~\ref{PSNspaceJO}, left panel) 
and we can determine them separately quite well.\\
Other correlations between the main parameters have not been found to be significant.
This is shown in Fig.~\ref{spaceALL} for SN-NGC6412 with added velocity prior.

\begin{figure}
\begin{center}
\includegraphics[width=\columnwidth]{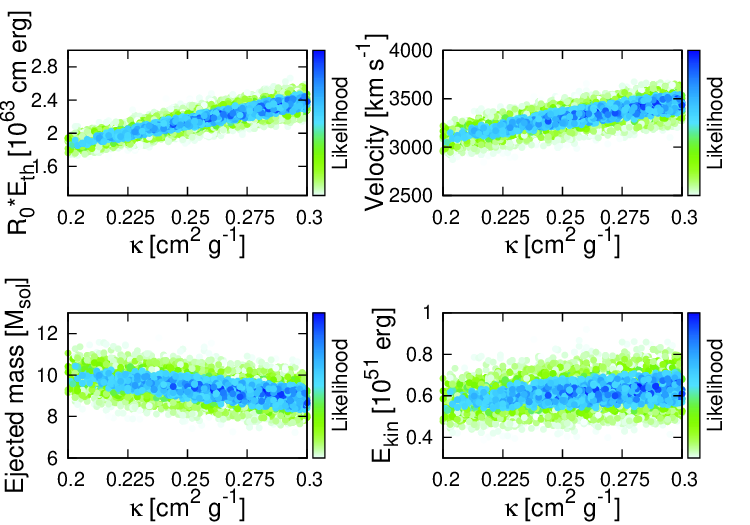} 
\caption{Opacity ($\kappa$) correlations for SN 2005cs. Within this range, it does not affect the parameters significantly. $s=0$.} 
\label{kapCOR}
\end{center}
\end{figure}

\citet{Nagy-Vinko-LC} showed that the opacity ($\kappa$) also correlates with $M_\text{ej}$.
We also sampled the opacity ($\kappa$) for additional correlations. A higher opacity increases both the velocity ($v$) and $R_0 \cdot E_\text{th}$, 
but $M_\text{ej}$ is the most affected. 
Increasing $\kappa$ substantially decreases $M_\text{ej}$, while $E_\text{kin}$ is also changing, although less significantly; (Fig.~\ref{kapCOR}).
This may cause a higher $M_\text{ej}$ while $E_\text{kin}$ remains about the same.
This may be the cause for higher mass estimations from hydrocodes, while other methods imply lower masses (see the Appendix) 
However care should be taken with the interpretation, as our model uses constant opacities, which is an approximation.\\
Nevertheless these correlations are not so significant and only increase the uncertainties of the parameters slightly.

\begin{figure}
\begin{center}
\includegraphics[width=\columnwidth]{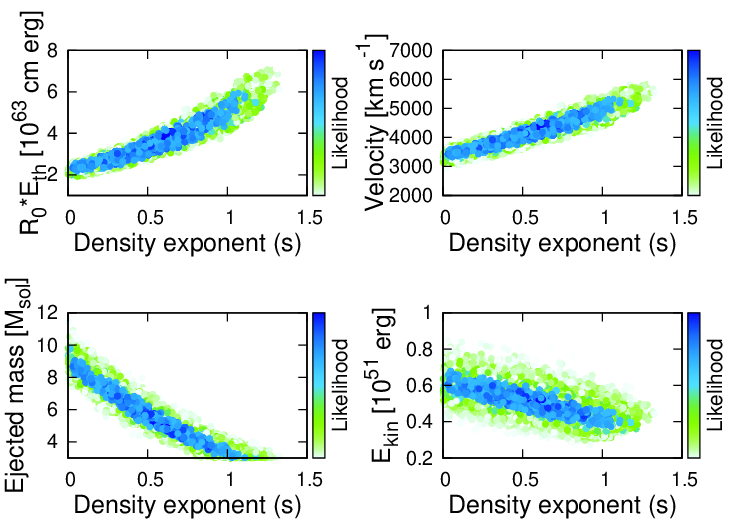} 
\caption{power-law density profile exponent correlations for SN 2005cs, $\kappa=[0.2:0.3]$ cm$^2$ g$^{-1}$.
Larger power-law density profile exponent values gives unrealistic results e.g very low mass and too high velocities.} 
\label{kapCOR2}
\end{center}
\end{figure}

Increasing the value of the power-law density profile exponent ($s$) 
significantly decreases $M_\text{ej}$ and also decreases $E_\text{kin}$, although less significantly, while 
$R_0 \cdot E_\text{th}$ increases. This is shown in Fig.~\ref{kapCOR2}.
These are significant correlations, so we adopt the $s=0$ value, because it gives the values consistent with the literature and the hydrocodes.
Larger power-law exponents give unrealistic results, like very low mass and too high velocities.
This confirms \citet{Nagy-Vinko-LC} where this statement was also tested. 
Correlation between $s$ and $\kappa$ have not been found in this parameter region.

Sampling $t_0$ and $d$ may also be possible, however doing so gives high uncertainties. 
Uncertainties in the date of plateau end and $t_0$ mostly increase the uncertainty of $M_\text{ej}$, 
but other parameters also become more uncertain,
although less significantly. Clearly it is more meaningful to have them determined independently.\\

The best values and the uncertainties of the fitted parameters for SN-NGC6412 are shown in Table~\ref{PSN_fitvalue}. 
The mean values, uncertainties and comparison values from the literature for the other SNe are shown 
in Tables $6 - 9$ in the Appendix
(for SN~1999em, SN~2004et, SN~2005cs, and SN~2012aw respectively).

\begin{figure}
\begin{center}
\includegraphics[width=\columnwidth]{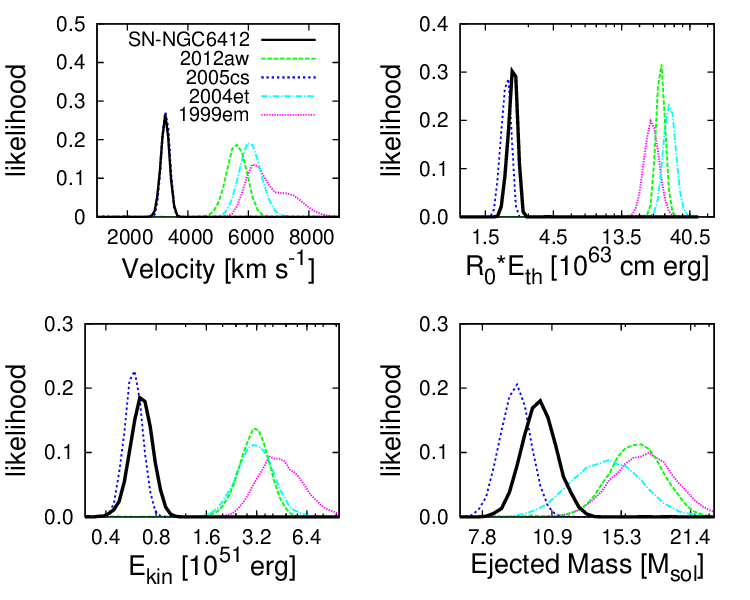} 
\caption{Parameter marginalization comparison, $\kappa=[0.2:0.3]$ cm$^2$ g$^{-1}$, $s=0$. (1999em: BB fit used, see text)} 
\label{vALL}
\end{center}
\end{figure}

For SN-NGC6412 the fitting gives $R_0 = 20-220 \cdot 10^{11} \;\text{cm} = 28-315 \; R_{\odot}$, $M_\text{ej} = 8.5 - 11.5 \; M_{\odot}$, 
the energies $0.5-0.8 \;\text{foe}$, $v_{exp} = 3300$ km s$^{-1}$. The initial nickel mass of SN-NGC6412 was $1.55_{-0.70}^{+0.75} \cdot 10^{-3} M_{\odot}$.
These are very similar to the values of SN  2005cs. 
Parameter comparison of the SNe is shown in (Fig.~\ref{vALL}). 
These results suggest that the progenitor of SN-NGC6412 had a moderate mass, and rather small radius, 
low velocities and energies along with very low nickel mass, similar to other LL SNe (\citealt{Pastorello-sublum}; \citealt{spiro-sublum}).
\citet{hamuy-parameters} showed correlation between energy and the nickel mass: larger nickel mass implies larger energy, 
which is consistent with our findings.\\

\begin{table*}
\begin{center}
\begin{tabular}{ l c c c c }
\hline
{\bf Parameter} & {\bf $2\sigma$ } & {\bf $1\sigma$ }  & {\bf $2\sigma$ } & {\bf $1\sigma$ } \\
\hline
$R_0$ [$10^{11} \;\text{cm}$]           & $176_{-155}^{+65}$      & $176_{-152}^{+55}$     & $91_{-70}^{+119}$        & $91_{-61}^{+110}$    \\
$M_\text{ej}$ [$M_{\odot}$]             & $9.90_{-1.34}^{+1.22}$  & $9.90_{-0.68}^{+0.42}$ & $9.89_{-1.00}^{+2.10}$   & $9.89_{-0.31}^{+0.83}$    \\
$E_\text{kin}$ [$10^{51} \;\text{erg}$] & $0.67_{-0.19}^{+0.18}$  & $0.67_{-0.07}^{+0.06}$ & $0.65_{-0.18}^{+0.19}$   & $0.65_{-0.06}^{+0.08}$    \\
$E_\text{th}$ [$10^{51} \;\text{erg}$]  & $0.15_{-0.05}^{+0.97}$  & $0.15_{-0.04}^{+0.83}$ & $0.28_{-0.17}^{+0.77}$   & $0.28_{-0.17}^{+0.70}$    \\
%$E_0$ [$10^{51} \;\text{erg}$]          & 0.82                    & 0.82                   & 0.93                     & 0.93        \\
$E_\text{th} R_0$ [$10^{62}$ erg cm]    & $26.3_{-3.1}^{+1.3}$    & $26.3_{-0.2}^{+0.2}$   & $25.6_{-5.0}^{+4.1}$     & $25.6_{-2.5}^{+1.0}$    \\
$v_\text{exp}$ [km s$^{-1}$]            & $3367_{-382}^{+181}$    & $3367_{-17}^{+17}$     & $3332_{-347}^{+216}$     & $3332_{-17}^{+17}$         \\ 
$M_\text{Ni}$ [$0.001 M_{\odot}$]       & $1.55_{-0.70}^{+0.75}$  & $1.55_{-0.70}^{+0.75}$ & $1.55_{-0.70}^{+0.75}$   & $1.55_{-0.70}^{+0.75}$ \\
$\kappa$ [cm$^2$ g$^{-1}$]              & 0.3                     & 0.3                    & 0.2:0.3   & 0.2:0.3         \\ 
\hline
\end{tabular}
\caption{SN-NGC6412 LC fit values. Power-law density profile exponent $s=0$.
The $2\sigma$ is for the 95\% confidence interval, and $1\sigma$ is for the 67\%.
The first two column is values with fixed opacity ($\kappa$), while in the second two column the $\kappa$ is sampled as well.
}
\label{PSN_fitvalue}
\end{center}
\end{table*}

\section{Summary}

We have made a photometric monitoring of SN-NGC6412 with two telescopes. It turned out that
SN-NGC6412 is a sub-luminous type II-P SN with low nickel mass very similar to the SN 2005cs.\\
There is only one spectrum of SN-NGC6412 available by \citet{toma-spectra}.
Fitting the spectrum with {\tt Syn++} reveals the presence of a strong \ion{He}{II} line, but quite weak H lines and low photospheric velocity.
\ion{He}{I} and \ion{N}{III} lines also present in the spectrum.
The early photospheric velocity and the expansion velocity are also very similar to SN 2005cs.\\
We fitted diluted blackbody radiation to the light curves in different photometric bands throughout the observed phases. 
The inferred photometric radii are smaller then those of regular II-P SNe, which also suggests low expansion velocities.\\

We modeled the bolometric light curve of SN-NGC6124 with our new fitting code that uses the MCMC method to find 
the most probable fitting parameters together with their exact uncertainties, and examine the correlations between the parameters.
With this we can reliably determinate the limits of our inferred parameters.\\
Besides the already known correlations between $M_\text{ej}$, $E_\text{kin}$, $R_0$ and $E_\text{th}$, 
we find that the optical opacity ($\kappa$) correlates with practically all other parameters, but the correlation is weak. 
$R_0$ and $E_\text{th}$ cannot be separated, and only their product can be determined from the fitting.
The correlation between $M_\text{ej}$, $E_\text{kin}$ (and $\kappa$) is weak, and both parameters can be determinated separately very well.
The power-law density profile exponent ($s$), however, shows significant correlations with the other parameters. 
We confirm that $s=0$ (constant density) results in the most realistic parameter values.\\

The most probable fitting parameters for SN-NGC6412 with 0.95 level confidence errors are the following: 
initial radius $R_0 = 91_{-70}^{+119} \cdot 10^{11} \;\text{cm}$, 
ejecta mass $M_\text{ej} = 9.89_{-1.00}^{+2.10} \; M_{\odot}$, 
kinetic energy $E_\text{kin} = 0.65_{-0.18}^{+0.19} \;\text{foe}$, expansion velocity $v_{\mbox{exp}} = 3332_{-347}^{+216}$ km s$^{-1}$. 
The initial nickel mass of SN-NGC6412 was $1.55_{-0.70}^{+0.75} \cdot 10^{-3} M_{\odot}$.
These values are very similar to those of SN 2005cs.\\
The inferred physical parameters of the ejecta suggest that SN-NGC6412 most probably arose from a moderate-mass progenitor just as SN~2005cs did.
This may give further support to the hypothesis that most of such low luminosity 
Type II-P SNe are due to the core collapse of moderate-mass (8--10 $M_{\odot}$) red supergiant (RSG) stars.
They are good candidates for the electron capture SNe with O-Ne-Mg core.
Obviously our present dataset is insufficient to decide whether SN-NGC6412 had such a core.
The family of LL SNe is still mysterious. Further studies are recommended.
\\

We also note that, despite the unfortunate fact that the brightness change of 
SN-NGC6412 has been not sampled adequately, which would give very high uncertainties,
by correctly including independent observational evidence in the analysis, namely radial velocity as an a priori probability distribution, 
the parameters have become quite well constrained.
This gives hope that the analysis of other similarly poorly sampled supernovae can also be attempted with the same methodology.

%\clearpage

\section*{Acknowledgments}
This work is part of the project
``Transient  Astrophysical  Objects``  GINOP  2.3.2-15-2016-00033 of the National Research, Development and
Innovation  Office  (NKFIH),  Hungary,  funded  by  the European  Union. \\

L. M. was supported by the Premium Postdoctoral Research Program of the Hungarian Academy of Sciences.
The research leading to these results has received funding from the LP2018-7 Lend\"ulet grant of the Hungarian Academy of Sciences.

Funding for the Sloan Digital Sky Survey IV has been provided by the Alfred P. Sloan Foundation, the U.S. Department of Energy Office of Science, and the Participating Institutions. SDSS-IV acknowledges
support and resources from the Center for High-Performance Computing at
the University of Utah. The SDSS web site is www.sdss.org.

\section*{Data availability}

The data underlying this article are available in
%\url{https://github.com/Hydralisk24/Science/tree/master/SN-LC-MCMC}
\url{https://github.com/Hydralisk24/Science}

\clearpage

\section*{Appendix}

In the Appendix we present a more detailed description about our fitting code, then show the
results of fitting the LCs of several other well-known Type II-P SNe, 
and compare them with the published parameter values for these SNe that were derived using various methods. \\

First, the numerical routines within the modeling part have been improved in order to make the running time as short as possible. 
This was an essential step, as the MCMC method needs to run the code hundred thousands or millions of times to give sensible results.
The algorithm which determines the ionization zone was also optimized.

The light curve is computed from three main components: $L$, $L_\text{ion}$, 
and $L_\text{pos}$ (see \citealt{Nagy-LC}; \citealt{Nagy-Vinko-LC} for more details).
$L$ is fully numerical at first, and computed with a fourth order Runge-Kutta method 
(for the differential equation see Eq.\/~(12) of \citealt{Nagy-LC}):

\begin{equation}
\frac{d \phi(t)}{dt} \tau_{Ni} = \frac{R(t)}{R_0 \cdot x_\text{i}^3} \cdot ( p_1 \zeta(t) - p_2 x_\text{i} \cdot \phi(t) - 2 \tau_{Ni} \cdot \phi(t) \cdot \frac{R_0}{R(t)} \frac{dx_\text{i}}{dt})
\end{equation}

$\zeta(t)$ is the energy coming from the $^{56}$Co and $^{56}$Ni decay, and $\tau_{Ni}$ is the decay time of the nickel. 
$\tau_d$ is the diffusion timescale (\citealt{Arnett-LC1}),  $p_1$ and $p_2$ are constants (see \citealt{Nagy-LC}).

\begin{equation}
L+L_\text{ion} = x_\text{i} \cdot \frac{\phi(t) E_\text{th}(0)}{\tau_d} \cdot (1 - e^{-A_g/t^2}) + 4 \pi r_\text{i}^2 Q \cdot \rho(x_\text{i},t) \cdot \frac{dr_\text{i}}{dt}
\label{NagyEtal2}
\end{equation}

%\begin{equation}
%L_\text{poz} = M_\text{Ni} \cdot (1 - e^{-A_g \frac{\kappa_{\text{poz}}}{\kappa_{\gamma}}/t^2}) \cdot (E_1 + E_2 \cdot (1 - e^{-A_g/t^2})) \cdot e^{-t/\tau_{Co}} - e^{-t/\tau_{Ni}}
%\end{equation}

\noindent where $Q$ is the recombination energy per unit mass, $\rho$ is the density.
$L_\text{ion}$ (the luminosity from ionization/recombination, 
$L+L_\text{ion}$ is equal with Eq.\.~(15) of \citet{Nagy-LC}. Eq.~(\ref{NagyEtal2}) is semi-analytic: the $dr_\text{i}$ component (layer width of the ionization zone) 
is calculated numerically, within the fourth order Runge-Kutta method. 
The ionization zone ($x_\text{i}$) is computed in every Runge-Kutta step, so $dr_\text{i}$ becomes far more accurate in this way.
$L_\text{pos}$ (luminosity from positrons, \citealt{Woosley-leak}; \citealt{Seitenzahl-leak}) is fully analytic.

The analytic nebular phase can be fitted separately. This be can be done because there are only two parameters that describe the nebular phase:
the nickel mass $M_\text{Ni}$ and the effective gamma-ray trapping $T_0$ (\citealt{CV-gammaleak}). 
We use the parameter $A_g = T_0^{2}$ instead (\citealt{Chatzopoulos-gammaleak}). Because this part is analytical, it is fast. 
From this, the set of best parameter pairs ($A_g$, $M_\text{Ni}$) is determined in the form of a 
function $M_\text{Ni}(A_g)$ which describes what nickel mass fits the nebular phase for various values of $A_g$.

$A_g$ has the following form (\citealt{CV-gammaleak}):
\begin{equation}
A_g = T_0^{2} =  \frac{\kappa_{\gamma} \cdotp M_\text{ej}}{4 \cdotp \pi \cdotp f \cdotp v^2},
\end{equation}
where $\kappa_{\gamma}$ is the gamma-ray opacity, $f$ is a geometric factor, $g_1$ is a geometric integral (see \citealt{Nagy-Vinko-LC} for more information), and
\begin{equation}
v^2 =  \frac{2 \cdotp E_\text{kin} \cdotp f}{g_1 \cdotp M_\text{ej}}
\label{v}
\end{equation}

So $E_\text{kin}$, and $M_\text{ej}$ ($\kappa_{\gamma}$ is not sampled) determine $A_g$, then $M_\text{Ni}$ 
can be inferred from $M_\text{Ni}(A_g)$ function to ensure that the nebular phase is fitted as closely as possible.

The physical equations and methods remained identical with those used in LC2.2. 
Also, the upgraded version has less numerical instabilities.
Because of this, the outputs of the two models -- the original LC2.2 and the new LC3.2 -- are essentially the same.\\

The parameters in our case are the initial radius ($R_0$), the ejected mass ($M_\text{ej}$), 
and the energies (total explosion energy: $E_0 = E_\text{kin} + E_\text{th}$, kinetic: $E_\text{kin}$, thermal: $E_\text{th}$).

All of them are set as scale parameters.
We adopted uniform a priori pdf-s \textsl{in their logarithms}, which corresponds to a regularized Jeffreys prior.
In fact, we directly sampled their logarithms in the algorithm rather than the parameters themselves.

The reported parameter values are to the mode of the joint posterior, corresponding to the best fitting solution.
The upper and lower uncertainty limits are derived from the 2 $\sigma$ (0.95 level) confidence intervals of the marginalized pdf
around the best solution.

In the MCMC the likelihood correlates with the sample number, so this simplifies the calculation, as only the sample number needs to be plotted, 
and the sample mean, standard deviation, measure of correlations and confidence intervals can be computed by simple summation over the chain elements.\\

As a result, we find good match between our results and those of others. Details are given below.

\begin{figure*}
\begin{center}
\includegraphics[width=\columnwidth]{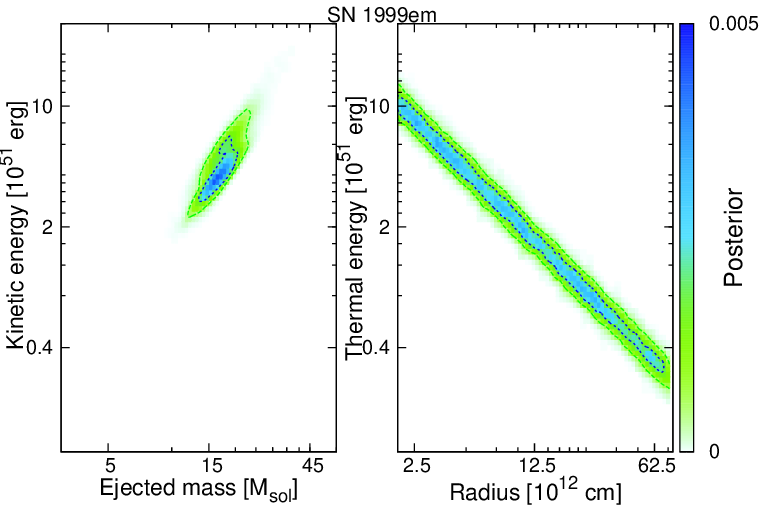} 
\includegraphics[width=\columnwidth]{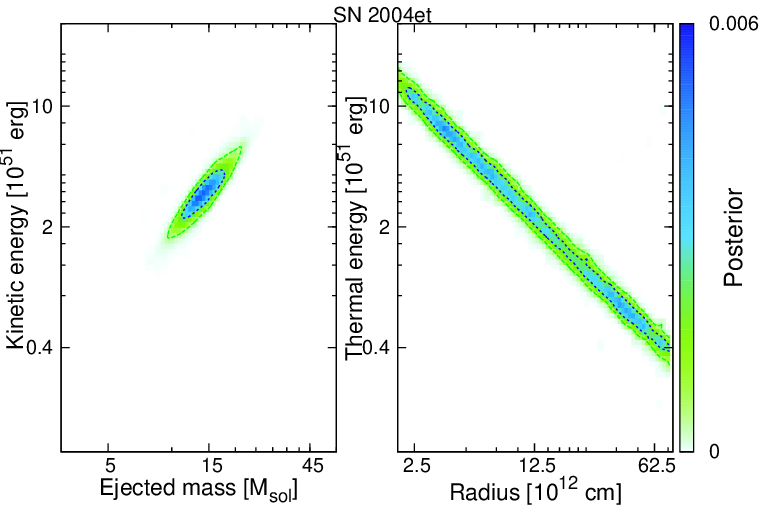} 
\includegraphics[width=\columnwidth]{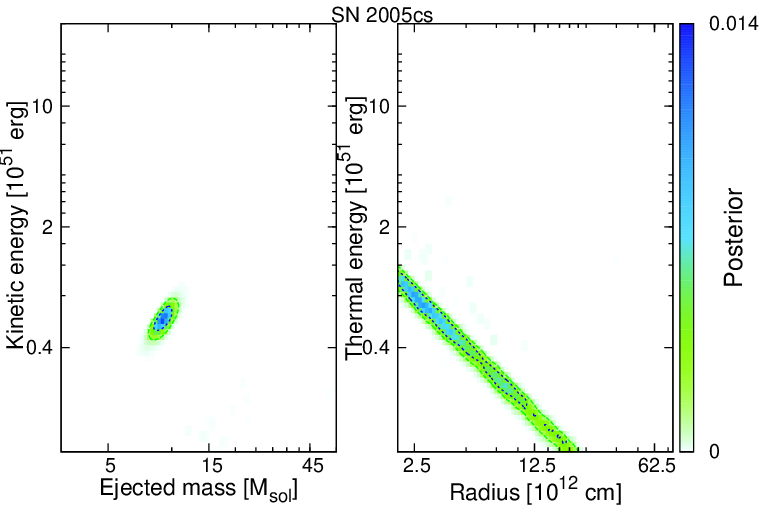} 
\includegraphics[width=\columnwidth]{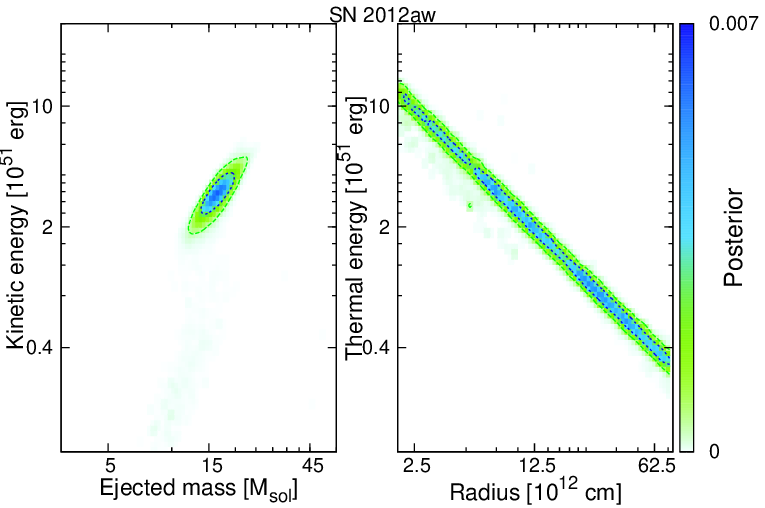} 
\caption{Other SNe probability distributions of the joint parameters, upper left: SN 1999em (BB fit; see text), upper right: SN 2004et, bottom left: SN 2005cs, bottom right: SN 2012aw.
$s=0$, $\kappa=[0.2:0.3]$ cm$^2$ g$^{-1}$. 
The colors encode the likelihood as the goodness of the fits.
The contours show the confidence interval of 67\% and 95\%.} 
\label{whole_space}
\end{center}
\end{figure*}

\begin{figure*}
\begin{center}
\includegraphics[width=5.0in]{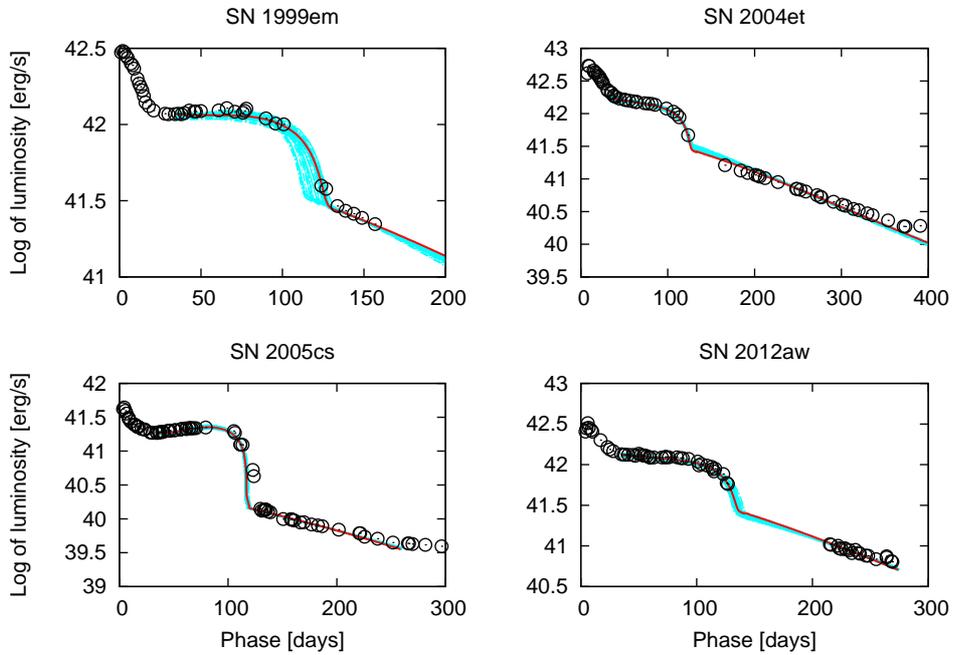} 
\caption{Model fits for the comparison SNe. The red line is the best fit. $s=0$, $\kappa=[0.2:0.3]$ cm$^2$ g$^{-1}$. (1999em: BB fit used; see text)} 
\label{other_best}
\end{center}
\end{figure*}

The posterior distribution and best fits for the reference SNe are shown in (Fig.~\ref{whole_space}) and in (Fig.~\ref{other_best}).
The values in Table 6 -- 9. 
%~\ref{1999em_fitvalue}, \ref{2004et_fitvalue}, \ref{2005cs_fitvalue}, and \ref{2012aw_fitvalue}.
Because of the correlations between various parameters, and the usage of various method gives a slightly different values in the literature.
The radius has a very strong correlation with the thermal energy giving very high uncertainty.
The mass values are more or less consistent.
The advantage of an MCMC sampling over simple optimization methods is that it also allows the consistent assesment of the parameter uncertainties. 
Considering this, we are in good agreement with previous results by \citet{Nagy-Vinko-LC} and others.
The energy median is somewhat larger, but the literature values are within our lower error limits. 
The other parameters are in very good agreement with the literature.

There are estimation formulae (\citealt{Litvinova-formulas}; \citealt{Nadyozhin-formulas}) to derive ejected mass, 
radius and energy from the light-curve shape without modeling it.
This is often used in literature, but may have large uncertainties. 
%This method labeled as 'formulae' method in the Tables.
The values inferred this method appear in the Tables in columns labeled as 'formulae'.
Note: $v_\text{exp}$ is not identical with photometric velocities, see \citet{Nagy-Vinko-LC}.

There are existing pre-explosion Hubble Space Telescope (HST) images which give independent observation values for the SNe parameters:
2005cs: $M_0 = 6-13 \; M_{\odot}$ (\citealt{Maund-05csrefHST}; \citealt{Li-05csrefHST}; \citealt{Eldridge-05csrefHST}).
2004et: Mass before explosion: $M_0 = 8-14 \; M_{\odot}$ (\citealt{Smartt-04etrefHST}).
2012aw: $M_0 = 17-18 \; M_{\odot}$ (\citealt{Vandyk-12awrefHST}), 
and $M_\text{ZAMS} = 14-26 \; M_{\odot}$ and $R_0 > 350 \cdot 10^{11}$ cm (\citealt{Fraser-12awrefHST}).\\

Our code presented in this paper can be downloaded from:\\
\url{https://github.com/Hydralisk24/Science/tree/master/SN-LC-MCMC}\\
Contact: jagerz24@gmail.com

\begin{table}
\begin{center}
\begin{tabular}{ l c c c c c c }
\hline
{\bf Parameter} & {\bf LC3.2 BB }$2\sigma$ & {\bf LC3.2 RJ }$2\sigma$  & {\bf Bose } & {\bf Utrobin } & {\bf Elmhandi } & {\bf Baklanov }\\
%                &  {\bf UBVRI }       &  {\bf UBVRI }        &  &  &  & \\
%{\bf Parameter} & {\bf LC3 } & {\bf LC3 } & {\bf LC3 alt} & {\bf \citealt{bose-2012aw} } & {\bf \citealt{Utrobin-99emref} } & {\bf \citealt{Elmhamdi-99emref} } & {\bf \citealt{Baklanov-99emref} }\\
\hline
Method               & semi- & semi- & semi- & hydro   & formulae  & formulae \\
                     & analytic  & analytic  & analytic  &  &  &  \\

\hline
$R_0$ [$10^{11}$ cm]                 & $659^{-639}_{+37}$      & $52^{-32}_{+645}$       & 280 $\pm$ 38 & 350 $\pm$ 140 & 95 $\pm$ 11 & 508 $\pm$ 193\\
$M_\text{ej}$ [$M_{\odot}$]          & $19.10^{-6.18}_{+3.57}$ & $13.16^{-2.04}_{+4.94}$ & 11 $\pm$ 3   & -             & 10--11      & 16.5 $\pm$ 1.5\\
$M_\text{0}$ [$M_{\odot}$]           & -                       & -                       & -            & 19 $\pm$ 1.2  & -           & - \\
$E_\text{kin}$ [$10^{51}$ erg]       & $4.21^{-1.55}_{+4.10}$  & $2.52^{-0.78}_{+2.19}$  & -            & -             & -      & - \\
$E_\text{th}$ [$10^{51}$ erg]        & $0.32^{-0.05}_{+9.98}$  & $3.47^{-3.21}_{+4.85}$  & -            & -             & -      & - \\
$E_0$ [$10^{51}$ erg]                & -                       & -                       & 0.5--0.9     & 1.3           & 0.5--1 & 0.7--1 \\
$E_\text{th} R_0$                    & $210_{-29}^{+63}$       & $179_{-36}^{+24}$       & -            & -             & -      & -  \\
.[$10^{62}$ erg cm]                  &                         &                         &              &               &        &    \\
$v_\text{exp}$ [km s$^{-1}$]         & $6078_{-455}^{+1865}$   & $5658_{-646}^{+1025}$   & -            & *             & -      & -  \\ 
$M_\text{Ni}$ [$M_{\odot}$]          & 0.06                    & 0.04                    & -            & 0.036         & 0.02   & -  \\
$\kappa$ [cm$^2$ g$^{-1}$]           & 0.2:0.3                 & 0.2:0.3                 & -            & -             & -      & -  \\

\hline
\end{tabular}
\caption{SN 1999em fit and literature values. Mass before explosion: $M_0$, Full energy: $E_0$.
Bose: \citealt{bose-2012aw}; Utrobin: \citealt{Utrobin-99emref}; Elmhandi: \citealt{Elmhamdi-99emref}; Baklanov: \citealt{Baklanov-99emref} 
*: Not identical velocity with our velocities. $s=0$.}
\label{1999em_fitvalue}
\end{center}
\end{table}

\begin{table*}
\begin{center}
\begin{tabular}{ l c c c c c c c }
\hline
{\bf Parameter} & {\bf LC3.2 }$2\sigma$ & {\bf LC3.2 }$2\sigma$  & {\bf Nagy,} & {\bf Sahu } & {\bf Bose } & {\bf Utrobin } & {\bf Misra }\\
                &            &            & {\bf Vinko} &             &             &                &             \\
\hline
%{\bf Parameter} & {\bf LC3 } & {\bf LC3 } & {\bf LC2 } & {\bf \citealt{sahu-2004et} } & {\bf \citealt{bose-2012aw} } & {\bf \citealt{Utrobin-04etref} } & {\bf \citealt{Misra-04etref} }\\
Method               & semi-     & semi-     & semi-     & formulae & formulae & hydro & formulae \\
                     & analytic  & analytic  & analytic  &  &  &  & \\
\hline
$R_0$ [$10^{11}$ cm]             & $306_{-284}^{+424}$     & $26_{-2}^{+703}$        & 420  & -               & 414 $\pm$ 63  & 1050 $\pm$ 98 & -    \\
$M_\text{ej}$ [$M_{\odot}$]      & $12.99_{-4.11}^{+2.59}$ & $17.76_{-7.43}^{+1.76}$ & 11.0 & 15 $\pm$ 5      & 9 $\pm$ 2     & -             & 12 $\pm$ 4 \\
$M_\text{0}$ [$M_{\odot}$]       & -                       & -                       & -    & -               & -             & 24.5 $\pm$ 1  & - \\
$E_\text{kin}$ [$10^{51}$ erg]   & $3.14_{-1.40}^{+1.56}$  & $3.85_{-1.85}^{+1.20}$  & 1.35 & -               & -             & -             & - \\
$E_\text{th}$ [$10^{51}$ erg]    & $1.02_{-0.63}^{+10.1}$  & $11.7_{-11.3}^{+0.19}$  & 0.60 & -               & -             & -             & - \\
$E_0$ [$10^{51}$ erg]            & -                       & -                       & 1.95 & 1.24 $\pm$ 0.34 & 0.6 $\pm$ 0.2 & 2.3 $\pm$ 0.3 & 1.96 $\pm$ 0.25 \\
$E_\text{th} R_0$                & $312_{-84}^{+32}$       & $307_{-64}^{+38}$       & -    & - & -  & - & - \\
.[$10^{62}$ erg cm]              &                         &                         &      &   &    &   &   \\
$v_\text{exp}$ [km s$^{-1}$]     & $6361_{-1052}^{+322}$   & $6029_{-720}^{+654}$    &4250& -  & - &*& -  \\
$M_\text{Ni}$ [$M_{\odot}$]      & 0.06                    & 0.06      & 0.06 & 0.06       & -       & 0.068     & 0.06      \\
$\kappa$ [cm$^2$ g$^{-1}$]       & 0.3                   & 0.2:0.3   & 0.3  & -          & -       & -         & -         \\

\hline
\end{tabular}
\caption{SN 2004et LC fit and literature values. Mass before explosion: $M_0$, Full energy: $E_0$.
Nagy, Vinko: \citealt{Nagy-Vinko-LC}; Sahu: \citealt{sahu-2004et}; Bose: \citealt{bose-2012aw}; Utrobin: \citealt{Utrobin-04etref}; 
Misra: \citealt{Misra-04etref}. *: Not identical velocity with our velocities. $s=0$.}
\label{2004et_fitvalue}
\end{center}
\end{table*}

\begin{table*}
\begin{center}
\begin{tabular}{ l c c c c c c c }
\hline
{\bf Parameter} & {\bf LC3.2 }$2\sigma$ & {\bf LC3.2 }$2\sigma$  & {\bf Nagy } & {\bf Utrobin } & {\bf Pastorello } & {\bf Takats } & {\bf Spiro }\\
%{\bf Parameter} & {\bf LC3 } & {\bf LC3 } & {\bf LC2 } & {\bf \citealt{Utrobin-05csref} } & {\bf \citealt{pastorello-2005cs} } & {\bf \citealt{Takats-05csref} }\\
\hline
Method               & semi-     & semi-     & semi-     & hydro & semi-    & formulae & hydro\\
                     & analytic  & analytic  & analytic  &       & analytic &  & \\
\hline
$R_0$ [$10^{11}$ cm]             & $218_{-198}^{+218}$    & $33_{-13}^{+158}$        & 120        & 420 $\pm$ 98 & 70         & 180 $\pm$ 126 & 250 \\
$M_\text{ej}$ [$M_{\odot}$]      & $8.84_{-1.19}^{+1.10}$ & $8.80_{-0.86}^{+1.52}$   & 8.00       & -            & 11 $\pm$ 3 & 9.7 $\pm$ 5.4 & 9.5 \\
$M_\text{0}$ [$M_{\odot}$]       & -                      & -                        & -          & 17 $\pm$ 1   & -          & -             & - \\
$E_\text{kin}$ [$10^{51}$ erg]   & $0.62_{-0.14}^{+0.17}$ & $0.62_{-0.14}^{+0.11}$   & 0.32       &  -           &  -         & -             & - \\
$E_\text{th}$ [$10^{51}$ erg]    & $0.11_{-0.01}^{+0.94}$ & $0.71_{-0.60}^{+0.27}$   & 0.16       &  -           &  -         & -             & - \\
$E_0$ [$10^{51}$ erg]            & -                      & -                        & 0.48       & 0.41         & 0.3        & 0.23 $\pm$ 0.14 & 0.16 \\   
$E_\text{th} R_0$ [$10^{62}$ erg cm]    & $23.8_{-2.0}^{+0.7}$    & $23.5_{-5.2}^{+0.2}$   & -    & -            & -          & -             & - \\
$v_\text{exp}$ [km s$^{-1}$]            & $3441_{-279}^{+107}$    & $3444_{-459}^{+94}$    & 2580 & *            & *          & -             & * \\ 
$M_\text{Ni}$ [$10^{-3} M_{\odot}$]     & 2.8                     & 2.8                    & 2    & 8.2          & 3          & 3             & 6 $\pm$ 3 \\
$\kappa$ [cm$^2$ g$^{-1}$]       & 0.3                     & 0.2:0.3                & 0.3  & -            & -          & -             & - \\

\hline
\end{tabular}
\caption{SN 2005cs fit and literature values. Mass before explosion: $M_0$, Full energy: $E_0$.
Nagy: \citealt{Nagy-Vinko-LC}; Utrobin: \citealt{Utrobin-05csref}; Pastorello: \citealt{pastorello-2005cs}; 
Takats: \citealt{Takats-05csref}; Spiro: \citealt{spiro-sublum}. *: Not identical velocity with our velocities. $s=0$.}
\label{2005cs_fitvalue}
\end{center}
\end{table*}

\begin{table*}
\begin{center}
\begin{tabular}{ l c c c c c }
\hline
{\bf Parameter} & {\bf LC3.2 }$2\sigma$ & {\bf LC3.2 }$2\sigma$ & {\bf Nagy, Vinko } & {\bf Bose } & {\bf Dall'ora }\\
%{\bf Parameter} & {\bf LC3 } & {\bf LC3 } & {\bf LC2 } & {\bf \citealt{bose-2012aw} } & {\bf \citealt{Dallora-12awref} }\\
\hline
Method       & semi-analytic & semi-analytic & semi-analytic & formulae & radiation hydro \\
\hline
$R_0$ [$10^{11}$ cm]           & $579_{-559}^{+117}$     & $286_{-265}^{+476}$     & 295  & 337 $\pm$ 67 & 300 \\
$M_\text{ej}$ [$M_{\odot}$]      & $14.92_{-3.37}^{+2.51}$ & $14.54_{-1.62}^{+6.49}$ & 20.0 & 14 $\pm$ 5 & 20 \\
$E_\text{kin}$ [$10^{51}$ erg]   & $3.07_{-1.34}^{+1.00}$  & $2.98_{-0.98}^{+1.73}$  & 1.60 & -       & -  \\
$E_\text{th}$ [$10^{51}$ erg]    & $0.47_{-0.11}^{+11.4}$  & $0.92_{-0.58}^{+8.67}$  & 0.60 & -       & -  \\
$E_0$ [$10^{51}$ erg]            & -                       & -                       & 2.20 & 1-2     & 1.5 \\
$E_\text{th} R_0$ [$10^{62}$ erg cm]    & $271_{-55}^{+18}$       & $264_{-48}^{+25}$      &  -  &  - & -   \\
$v_\text{exp}$ [km s$^{-1}$]            & $5875_{-863}^{+434}$    & $5855_{-1123}^{+102}$  &3650& -  & * \\ 
$M_\text{Ni}$ [$M_{\odot}$]             & 0.06                    & 0.06                   & 0.06 & 0.06    & 0.06 \\
$\kappa$ [cm$^2$ g$^{-1}$]              & 0.3                     & 0.2:0.3                & 0.13 & -       & - \\

\hline
\end{tabular}
\caption{SN 2012aw fit and literature values. Mass before explosion: $M_0$, Full energy: $E_0$.
Nagy, Vinko: \citealt{Nagy-Vinko-LC}; Bose: \citealt{bose-2012aw}; Dall'ora: \citealt{Dallora-12awref}. 
*: Not identical velocity with our velocities. $s=0$.}
\label{2012aw_fitvalue}
\end{center}
\end{table*}

%%%%%%%%%%%% 1999em ALT van benne


\begin{thebibliography}{}



\bibitem[\protect\citeauthoryear{Albareti et al.}{ 2017}]{dr13} 
Albareti F.~D. et al., 2017, ApJS, 233, 25

\bibitem[\protect\citeauthoryear{Arnett}{ 1980}]{Arnett-LC1}
Arnett W. D., 1980, ApJ, 237, 541A

\bibitem[\protect\citeauthoryear{Arnett}{ 1982}]{Arnett-LC2}
Arnett W. D., 1982, ApJ, 253, 785A

\bibitem[\protect\citeauthoryear{Arnett \& Fu}{ 1989}]{Arnett-Fu-LC}
Arnett W. D., Fu A., 1989, ApJ, 340, 396

%\bibitem[\protect\citeauthoryear{Baade}{ 1926}]{baade-EPM}
%Baade W., 1926, Astr. Nachr., 228, 359

\bibitem[\protect\citeauthoryear{Baklanov et al.}{ 2005}]{Baklanov-99emref}
Baklanov P. V., Blinnikov S. I., Pavlyuk N. N., 2005, AstL, 31, 429B

\bibitem[\protect\citeauthoryear{Bayless et al.}{ 2013}]{Bayless-12awUV}
Bayless A. J. et al., 2013, ApJ, 764L, 13B
%Bayless, Amanda J.; Pritchard, Tyler A.; Roming, Peter W. A.; Kuin, Paul; Brown, Peter J.; Botticella, Maria Teresa; Dall'Ora, Massimo; 
%Frey, Lucille H.; Even, Wesley; Fryer, Chris L.; Maund, Justyn R.; Fraser, Morgan, 2013, ApJ, 764L, 13B

\bibitem[\protect\citeauthoryear{Benetti et al.}{ 2001}]{benetti-1997d}
Benetti S. et al., 2001, MNRAS, 322, 361B
%Benetti, S., Turatto, M., Balberg, S., Zampieri, L., Shapiro, S. L., Cappellaro, E., 
%Nomoto, K., Nakamura, T., Mazzali, P. A., Patat, F., 2001, MNRAS, 322, 361B

\bibitem[\protect\citeauthoryear{Blinnikov \& Popov}{ 1993}]{Blinn-Popov-LC}
Blinnikov S. \& Popov D. V., 1993, A\&A, 274, 775

\bibitem[\protect\citeauthoryear{Blondin \& Tonry}{ 2007}]{tonry-snid}
Blondin S., Tonry J. L., 2007, ApJ, 666, 1024B 

%\bibitem[\protect\citeauthoryear{Blondin \& Tonry}{ 2011}]{tonry-snid}
%Blondin, St馥hane; Tonry, John L., 2011, ascl.soft 07001B

\bibitem[\protect\citeauthoryear{Bose et al.}{ 2013}]{bose-2012aw}
Bose S. et al., 2013, MNRAS, 433, 1871B
%Bose, Subhash, Kumar, Brijesh, Sutaria, Firoza, Kumar, Brajesh, Roy, Rupak, Bhatt, V. K., 
%Pandey, S. B., Chandola, H. C., Sagar, Ram, Misra, Kuntal, Chakraborti, Sayan, 2013, MNRAS, 433, 1871B

\bibitem[\protect\citeauthoryear{Bottinelli et al.}{ 1984}]{Bottinelli-distance}
Bottinelli L., Gouguenheim L., Paturel G., de Vaucouleurs G., 1984, A\&AS, 56, 381B

\bibitem[\protect\citeauthoryear{Bottinelli et al.}{ 1986}]{Bottinelli-distance2}
Bottinelli L., Gouguenheim L., Paturel G., Teerikorpi P., 1986, A\&A, 156, 157B

\bibitem[\protect\citeauthoryear{Brown et al.}{ 2007}]{Brown-05csrefUV}
Brown P. J. et al., 2007, ApJ, 659, 1488B
%Brown, Peter J.; Dessart, Luc; Holland, Stephen T.; Immler, Stefan; Landsman, Wayne; Blondin, St馥hane; Blustin, Alexander J.; 
%Breeveld, Alice; Dewangan, Gulab C.; Gehrels, Neil; Hutchins, Robert B.; Kirshner, Robert P.; Mason, Keith O.; Mazzali, Paolo A.; 
%Milne, Peter; Modjaz, Maryam; Roming, Peter W. A., 2007, ApJ, 659, 1488B

%SN theory cikk, 8sunmas...
\bibitem[\protect\citeauthoryear{Burrows}{ 2013}]{Burrows-SNtheory}
Burrows A., 2013, RvMP, 85, 245B

\bibitem[\protect\citeauthoryear{Burrows et al.}{ 2019}]{Burrows-SNtheory-2}
Burrows A., Radice D., Vartanyan D., 2019, MNRAS, 485, 3153B

\bibitem[\protect\citeauthoryear{Chugai \& Utrobin}{ 2000}]{Chugai-ecapture}
Chugai N. N., \& Utrobin V. P., 2000, A\&A, 354, 557

\bibitem[\protect\citeauthoryear{Chatzopoulos et al.}{ 2012}]{Chatzopoulos-gammaleak}
Chatzopoulos E.; Wheeler J. C., Vinko J., 2012, ApJ, 746, 121C

\bibitem[\protect\citeauthoryear{Clocchiatti \& Wheeler}{ 1997}]{CV-gammaleak}
Clocchiatti A., Wheeler J. C., 1997, ApJ, 491, 375C

\bibitem[\protect\citeauthoryear{Dall'ora et al.}{ 2014}]{Dallora-12awref}
Dall'Ora M. et al., 2014, ApJ, 787, 139D
%Dall'Ora, M.; Botticella, M. T.; Pumo, M. L.; Zampieri, L.; Tomasella, L.; Pignata, G.; Bayless, A. J.; Pritchard, T. A.; Taubenberger, S.; 
%Kotak, R.; Inserra, C.; Della Valle, M.; Cappellaro, E.; Benetti, S.; Benitez, S.; Bufano, F.; Elias-Rosa, N.; Fraser, M.; Haislip, J.  B.; 
%Harutyunyan, A.; Howell, D. A.; Hsiao, E. Y.; Iijima, T.; Kankare, E.; Kuin, P.; Maund, J. R.; Morales-Garoffolo, A.; Morrell, N.; Munari, U.; 
%Ochner, P.; Pastorello, A.; Patat, F.; Phillips, M. M.; Reichart, D.; Roming, P. W. A.; Siviero, A.; Smartt, S. J.; Sollerman, J.; Taddia, F.; 
%Valenti, S.; Wright, D., 2014, ApJ, 787, 139D

\bibitem[\protect\citeauthoryear{Dessart \& Hillier}{ 2005}]{dess-hill-BB}
Dessart L., Hillier D. J., 2005, A\&A, 439, 671D

\bibitem[\protect\citeauthoryear{Dessart et al.}{ 2008}]{dessart-2005cs}
Dessart L. et al., 2008, ApJ, 675, 644D
%Dessart, Luc, Blondin, St馥hane, Brown, Peter J., Hicken, Malcolm, Hillier, D. John, Holland, Stephen T., 
%Immler, Stefan, Kirshner, Robert P., Milne, Peter, Modjaz, Maryam, Roming, Peter W. A., 2008, ApJ, 675, 644D

\bibitem[\protect\citeauthoryear{Eastman et al.}{ 1996}]{eastman-diluted}
Eastman R. G., Schmidt B. P., Kirshner R., 1996, ApJ 466, 911

\bibitem[\protect\citeauthoryear{Eldridge et al.}{ 2007}]{Eldridge-05csrefHST}
Eldridge J. J., Mattila S., Smartt S. J., 2007, MNRAS, 376, L52

\bibitem[\protect\citeauthoryear{Elmhamdi et al.}{ 2003}]{Elmhamdi-99emref}
Elmhamdi A. et al., 2003, MNRAS, 338, 939E
%Elmhamdi, Abouazza; Danziger, I. J.; Chugai, N.; Pastorello, A.; Turatto, M.; Cappellaro, E.; Altavilla, G.; Benetti, S.; 
%Patat, F.; Salvo, M., 2003, MNRAS, 338, 939E

\bibitem[\protect\citeauthoryear{Fisher et al.}{ 1997}]{fisher-syn}
Fisher A., Branch D., Nugent P., Baron E., 1997, ApJ, 481, L89

\bibitem[\protect\citeauthoryear{Fixsen et al.}{ 1996}]{fixsen-CMB}
Fixsen D. J., Cheng E. S., Gales J. M., Mather J. C., Shafer R. A., Wright E. L., 1996, ApJ, 473, 576F 

\bibitem[\protect\citeauthoryear{Fraser et al.}{ 2011}]{Fraser-2009md}
Fraser M. et al., 2011, MNRAS, 417., 1417F
%Fraser, M.; Ergon, M.; Eldridge, J. J.; Valenti, S.; Pastorello, A.; Sollerman, J.; Smartt, S. J.; Agnoletto, I.; Arcavi, I.; Benetti, S.; 
%Botticella, M.-T.; Bufano, F.; Campillay, A.; Crockett, R. M.; Gal-Yam, A.; Kankare, E.; Leloudas, G.; Maguire, K.; Mattila, S.; Maund, J. R.; 
%Salgado, F.; Stephens, A.; Taubenberger, S.; Turatto, M., 
%2011, MNRAS, 417., 1417F

\bibitem[\protect\citeauthoryear{Fraser et al.}{ 2012}]{Fraser-12awrefHST}
Fraser M. et al., 2012, ApJ, 759L, 13F
%Fraser, M.; Maund, J. R.; Smartt, S. J.; Botticella, M.-T.; Dall'Ora, M.; Inserra, C.; Tomasella, L.; Benetti, S.; Ciroi, S.; 
%Eldridge, J. J.; Ergon, M.; Kotak, R.; Mattila, S.; Ochner, P.; Pastorello, A.; Reilly, E.; Sollerman, J.; Stephens, A.; Taddia, F.; 
%Valenti, S., 2012, ApJ, 759L, 13F

\bibitem[\protect\citeauthoryear{Gal-Yam et al.}{ 2011}]{Gal-Yam2011}
Gal-Yam A. et al., 2011, ApJ, 736, 159

\bibitem[\protect\citeauthoryear{Gilks et al.}{ 1996}]{Gilks-mcmc}
Gilks W. R., Richardson S., Spiegelhalter D. J., Markov Chain Monte Carlo in Practice (Chapman \& Hall: London, 1996)

\bibitem[\protect\citeauthoryear{Gutierrez et al.}{ 2017}]{gutirez-apere}
Gutierrez C. P. et al., 2017, ApJ, 850, 89G
%Gutierrez, Claudia P.; Anderson, Joseph P.; Hamuy, Mario; et al., 2017, ApJ, 850, 89G

\bibitem[\protect\citeauthoryear{Hamuy}{ 2003}]{hamuy-parameters}
Hamuy M., 2003, ApJ, 582, 905H

\bibitem[\protect\citeauthoryear{Hamuy et al.}{ 2001}]{hamuy-diluted}
Hamuy M. et al. 2001, ApJ, 558, 615 (H01)
%Hamuy M., Pinto P. A., Maza J., et al.

\bibitem[\protect\citeauthoryear{Hamuy \& Pinto}{ 2002}]{hamuy-SCM}
Hamuy M., Pinto P. A., 2002, ApJ, 566L, 63H

\bibitem[\protect\citeauthoryear{Hastings}{ 1970}]{Hastings-mcmc}
Hastings W. K., 1970, Biometrika, 57, 97

\bibitem[\protect\citeauthoryear{Hatano et al.}{ 1999}]{hatano-syn}
Hatano K., Branch D., Fisher A., Millard J., Baron E., 1999, ApJS, 121, 233H
%Hatano, Kazuhito; Branch, David; Fisher, Adam; Millard, Jennifer; Baron, E., 1999, ApJS, 121, 233H

\bibitem[\protect\citeauthoryear{Jerkstrand et al.}{ 2017}]{Jerkstrand-EC/Fecore}
Jerkstrand A., Ertl T., Janka H.-T., M\"uller E., 2017, MNRAS, Vol. 88, 278

\bibitem[\protect\citeauthoryear{Jordi et al.}{ 2006}]{jordi-filter}
Jordi K., Grebel E. K., Ammon K., 2006, A\&A, 460, 339J

\bibitem[\protect\citeauthoryear{Kirshner \& Kwan}{ 1974}]{Kirshner-EPM}
Kirshner R. P., Kwan J., 1974, ApJ, 193, 27

\bibitem[\protect\citeauthoryear{Kitaura, Janka, Hillebrandt}{ 2006}]{Kitaura-ecapture}
Kitaura F. S., Janka H.-Th., Hillebrandt W., 2006, A\&A, 450, 345

\bibitem[\protect\citeauthoryear{Leonard et al.}{ 2002}]{leonard-1999em}
Leonard D. C. et al., 2002, PASP, 114, 35L
%Leonard, Douglas C., Filippenko, Alexei V., Gates, Elinor L., Li, Weidong, Eastman, Ronald G., Barth, Aaron J., Bus, Schelte J., 
%Chornock, Ryan, Coil, Alison L., Frink, Sabine, Grady, Carol A., Harris, Alan W., Malkan, Matthew A., Matheson, Thomas, 
%Quirrenbach, Andreas, Treffers, Richard R., 2002, PASP, 114, 35L

\bibitem[\protect\citeauthoryear{Leonard et al.}{ 2003}]{leonard-1999em-dist}
Leonard D. C., Kanbur S. M., Ngeow C. C., Tanvir N. R., 2003, ApJ, 594, 247.

\bibitem[\protect\citeauthoryear{Li et al.}{ 2006}]{Li-05csrefHST}
Li W. et al., 2006, ApJ, 641, 1060
%Li W., Van Dyk S. D., Filippenko A. V., et al. 2006, ApJ, 641, 1060

\bibitem[\protect\citeauthoryear{Lisakov et al.}{ 2018}]{Sergey-sublum}
Lisakov S. M., Dessart L., Hillier D. J., Waldman R., Livne E.,  2018, MNRAS, 473, 3863L
%Lisakov, Sergey M.; Dessart, Luc; Hillier, D. John; Waldman, Roni; Livne, Eli  2018, MNRAS, 473, 3863L

%formulas:
\bibitem[\protect\citeauthoryear{Litvinova \& Nadyozhin}{ 1985}]{Litvinova-formulas}
Litvinova Y., Nadyozhin D. K., 1985, SvA Lett., 11, 45

\bibitem[\protect\citeauthoryear{Lyman et al.}{ 2014}]{lyman-bc}
Lyman J. D., Bersier D., James P. A., 2014, MNRAS, 437, 3848L

\bibitem[\protect\citeauthoryear{Maguire et al.}{ 2010}]{Maguire-04etrefNIR}
Maguire K. et al., 2010, MNRAS, 404, 981M
%Maguire, K.; Di Carlo, E.; Smartt, S. J.; Pastorello, A.; Tsvetkov, D. Yu.; Benetti, S.; Spiro, S.; Arkharov, A. A.; Beccari, G.; 
%Botticella, M. T.; Cappellaro, E.; Cristallo, S.; Dolci, M.; Elias-Rosa, N.; Fiaschi, M.; Gorshanov, D.; Harutyunyan, A.; 
%Larionov, V. M.; Navasardyan, H.; Pietrinferni, A.; Raimondo, G.; di Rico, G.; Valenti, S.; Valentini, G.; Zampieri, L.,
%2010, MNRAS, 404, 981M

\bibitem[\protect\citeauthoryear{Mattila et al.}{ 2008}]{Mattila2008}
Mattila S., Smartt S. J., Eldridge J. J., Maund J. R., Crock-ett R. M., Danziger I. J., 2008, ApJ, 688, L91

\bibitem[\protect\citeauthoryear{Maund et al.}{ 2005}]{Maund-05csrefHST}
Maund J. R., Smartt S. J., Danziger I. J., 2005, MNRAS, 364, L33

\bibitem[\protect\citeauthoryear{Metropolis et al.}{ 1953}]{Metropolis-mcmc}
Metropolis N., Rosenbluth A. W., Rosenbluth M. N., Teller A. H., Teller E., 1953, J.Chem.Phys., 21, 1087

\bibitem[\protect\citeauthoryear{Misra et al.}{ 2007}]{Misra-04etref}
Misra K., Pooley D., Chandra P., Bhattacharya D., Ray A. K., Sagar R., Lewin W. H. G., 2007, MNRAS, 381, 280M
%Misra, Kuntal; Pooley, Dave; Chandra, Poonam; Bhattacharya, D.; Ray, Alak K.; Sagar, Ram; Lewin, Walter H. G., 2007, MNRAS, 381, 280M

\bibitem[\protect\citeauthoryear{Morozova et al.}{ 2015}]{Morozova-SNEC}
Morozova V. S. et al., 2015, ApJ, 814, 63
%Morozova, V. S., Piro, A. L., Renzo, M., Ott, C. D., et al., 2015, ApJ, 814, 63

\bibitem[\protect\citeauthoryear{Mould et al.}{ 2000}]{mould-virgoinfall}
Mould J. R. et al., 2000, ApJ, 529, 786M 
%Mould, Jeremy R.; Huchra, John P.; Freedman, Wendy L.; Kennicutt, Robert C., Jr.; Ferrarese, Laura; Ford, Holland C.; Gibson, Brad K.; 
%Graham, John A.; Hughes, Shaun M. G.; Illingworth, Garth D.; Kelson, Daniel D.; Macri, Lucas M.; Madore, Barry F.; 
%Sakai, Shoko; Sebo, Kim M.; Silbermann, Nancy A.; Stetson, Peter B., 2000, ApJ, 529, 786M 

%\bibitem[\protect\citeauthoryear{NED}{}]{ref-NED} %%%%%%%%%%%%%%%%%%%%%%%%%%%%%%%%%%%%%%%%!!!!!!!!!!!!!!!!!!!!!!!!!!!!!!!!!!!!!!!!!!!!!!!!!!!!!!!
%NASA/IPAC Extragalactic Database, https://ned.ipac.caltech.edu

\bibitem[\protect\citeauthoryear{Nadyozhin}{ 2003}]{Nadyozhin-formulas}
Nadyozhin D. K., 2003, MNRAS, 346, 97

\bibitem[\protect\citeauthoryear{Nagy et al.}{ 2014}]{Nagy-LC}
Nagy A. P., Ordasi A., Vinko J., Wheeler J. C., 2014, A\&A, 571, A77

\bibitem[\protect\citeauthoryear{Nagy \& Vinko}{ 2016}]{Nagy-Vinko-LC}
Nagy A. P., Vinko J., 2016, A\&A, 589, A53

\bibitem[\protect\citeauthoryear{Nagy}{ 2018}]{Nagy-Ibc}
Nagy A. P., 2018, ApJ, 862, 143N

\bibitem[\protect\citeauthoryear{Nakaoka et al.}{ 2018}]{tatsuya-massiveLL}
Nakaoka T. et al., 2018, ApJ, 859, 78N
%Nakaoka, Tatsuya; Kawabata, Koji S.; Maeda, Keiichi; Tanaka, Masaomi; Yamanaka, Masayuki; Moriya, Takashi J.;
%Tominaga, Nozomu; Morokuma, Tomoki; Takaki, Katsutoshi; Kawabata, Miho; Kawahara, Naoki; Itoh, Ryosuke;
%Shiki, Kensei; Mori, Hiroki; Hirochi, Jun; Abe, Taisei; Uemura, Makoto; Yoshida, Michitoshi; Akitaya, Hiroshi; Moritani, Yuki;
%Ueno, Issei; Urano, Takeshi; Isogai, Mizuki; Hanayama, Hidekazu; Nagayama, Takahiro,
%2018, ApJ, 859, 78N

\bibitem[\protect\citeauthoryear{Pastorello et al.}{ 2004}]{Pastorello-sublum}
Pastorello A. et al., 2004, MNRAS, 347, 74
%Pastorello, A.; Zampieri, L.; Turatto, M.; Cappellaro, E.; Meikle, W. P. S.; Benetti, S.; Branch, D.; Baron, E.;
%Patat, F.; Armstrong, M.; Altavilla, G.; Salvo, M.; Riello, M., 
%2004, MNRAS, 347, 74

\bibitem[\protect\citeauthoryear{Pastorello et al.}{ 2006}]{Pastorello2006}
Pastorello A. et al., 2006, MNRAS, 370, 1752P
%Pastorello, A.; Sauer, D.; Taubenberger, S.; Mazzali, P. A.; Nomoto, K.; Kawabata, K. S.; Benetti, S.; Elias-Rosa, N.; 
%Harutyunyan, A.; Navasardyan, H.; Zampieri, L.; Iijima, T.; Botticella, M. T.; di Rico, G.; Del Principe, M.; Dolci, M.; 
%Gagliardi, S.; Ragni, M.; Valentini, G., 2006, MNRAS, 370, 1752P

\bibitem[\protect\citeauthoryear{Pastorello et al.}{ 2009}]{pastorello-2005cs}
Pastorello A. et al., 2009, MNRAS, 394, 2266P
%Pastorello, A., Valenti, S., Zampieri, L., Navasardyan, H., Taubenberger, S., Smartt, S. J., Arkharov, A. A., 
%B雷nbantner, O., Barwig, H., Benetti, S., Birtwhistle, P., Botticella, M. T., Cappellaro, E., Del Principe, M., 
%di Mille, F., di Rico, G., Dolci, M., Elias-Rosa, N., Efimova, N. V., Fiedler, M., Harutyunyan, A., H寢lich, P. A., 
%Kloehr, W., Larionov, V. M., Lorenzi, V., Maund, J. R., Napoleone, N., Ragni, M., Richmond, M., Ries, C., Spiro, S., 
%Temporin, S., Turatto, M., Wheeler, J. C., 2009, MNRAS, 394, 2266P

\bibitem[\protect\citeauthoryear{Planck Collaboration}{ 2018}]{plank-H0}
Planck Collaboration, 2018, A\&A, in press, arXiv:1807.06209
%https://ui.adsabs.harvard.edu/abs/2018arXiv180706209P/abstract


\bibitem[\protect\citeauthoryear{Popov}{ 1993}]{Popov-LC}
Popov D. V., 1993, ApJ, 414, 712

\bibitem[\protect\citeauthoryear{Poznanski et al.}{ 2012}]{poznanski-extinction}
Poznanski D., Prochaska J. X., Bloom J. S., 2012, MNRAS, 426, 1465P
%Poznanski, Dovi, Prochaska, J. Xavier, Bloom, Joshua S., 2012b, MNRAS, 426, 1465P

\bibitem[\protect\citeauthoryear{Pumo et al.}{ 2017}]{Pumo-sublum}
Pumo M. L., Zampieri L., Spiro S., Pastorello A., Benetti S., Cappellaro E., Manico G., Turatto M., 2017, MNRAS, 464, 3013P

%\bibitem[\protect\citeauthoryear{Ron Arbour}]{discovery}
%Ron Arbour amateur astronomer, http://mstecker.com/pages/apparbour.htm, www.rochesterastronomy.org
%https://www.britastro.org/journal_item/6572
%{toma-spectra}!

\bibitem[\protect\citeauthoryear{Riess et al.}{ 2020}]{riess-H0}
Riess A. G., 2020, Nat. Rev. Phys. 2, 10

\bibitem[\protect\citeauthoryear{Roy et al.}{ 2011}]{Roy2011}
Roy R. et al., 2011, ApJ, 736, 76

\bibitem[\protect\citeauthoryear{Sahu et al.}{ 2006}]{sahu-2004et}
Sahu D. K., Anupama G. C., Srividya S., Muneer S., 2006, MNRAS, 372, 1315S

\bibitem[\protect\citeauthoryear{Schlafly \& Finkbeiner}{ 2011}]{Schlafly-ref-NED}
Schlafly E.F., Finkbeiner P. D., 2011,  ApJ 737, 103
%NASA/IPAC Extragalactic Database, https://ned.ipac.caltech.edu

\bibitem[\protect\citeauthoryear{Seitenzahl et al.}{ 2014}]{Seitenzahl-leak}
Seitenzahl I. R., Timmes F. X., Magkotsios G., 2014, ApJ, 792, 10S

\bibitem[\protect\citeauthoryear{Smartt et al.}{ 2009}]{Smartt-04etrefHST}
Smartt S. J., Eldridge J. J., Crockett R. M., Maund J. R., 2009, MNRAS, 395, 1409

\bibitem[\protect\citeauthoryear{Spiro et al.}{ 2014}]{spiro-sublum}
Spiro S. et al., 2014, MNRAS, 439, 2873S
%Spiro, S.; Pastorello, A.; Pumo, M. L.; Zampieri, L.; Turatto, M.; Smartt, S. J.; Benetti, S.; Cappellaro, E.; Valenti, S.;
%Agnoletto, I.; Altavilla, G.; Aoki, T.; Brocato, E.; Corsini, E. M.; Di Cianno, A.; Elias-Rosa, N.; Hamuy, M.; Enya, K.;
%Fiaschi, M.; Folatelli, G.; Desidera, S.; Harutyunyan, A.; Howell, D. A.; Kawka, A.; Kobayashi, Y.; Leibundgut, B.;
%Minezaki, T.; Navasardyan, H.; Nomoto, K.; Mattila, S.; Pietrinferni, A.; Pignata, G.; Raimondo, G.; Salvo, M.;
%Schmidt, B. P.; Sollerman, J.; Spyromilio, J.; Taubenberger, S.; Valentini, G.; Vennes, S.; Yoshii, Y.,
%2014, MNRAS, 439, 2873S

\bibitem[\protect\citeauthoryear{Takats \& Vinko}{ 2006}]{Takats-05csref}
Takats K., Vinko J., 2006, MNRAS, 372, 1735T
%https://ui.adsabs.harvard.edu/abs/2006MNRAS.372.1735T/abstract

\bibitem[\protect\citeauthoryear{Takats et al.}{ 2014}]{Takats2014}
Takats K. et al., 2014, MNRAS, 438, 368T
%https://ui.adsabs.harvard.edu/abs/2014MNRAS.438..368T/abstract

\bibitem[\protect\citeauthoryear{Thomas et al.}{ 2011}]{thomas-syn}
Thomas R. C., Nugent P. E., Meza J. C., 2011, PASP, 123, 237

\bibitem[\protect\citeauthoryear{Tomasella et al.}{ 2015}]{toma-spectra}
Tomasella L. et al., 2015, ATel, 7787, 1T
%Tomasella, L., Benetti, S., Cappellaro, E., Elias-Rosa, N., Ochner, P., Pastorello, A., Tartaglia, L., 
%Terreran, G., Turatto, M., 2015, ATel, 7787, 1T

\bibitem[\protect\citeauthoryear{Tully \& Fisher}{ 1988}]{Tully-Fisher}
Tully R. Brent, Fisher J. R., 1988, Cambridge University Press, cng, book T

%elektrom cepture es failed explosion ref
\bibitem[\protect\citeauthoryear{Turatto et al.}{ 1998}]{Turatto-ecapture}
Turatto M. et al., 1998, ApJ, 498, L129
%Turatto, M.; Mazzali, P. A.; Young, T. R.; Nomoto, K.; Iwamoto, K.; Benetti, S.; Cappellaro, E.; Danziger, I. J.; de Mello, D. F.;
%Phillips, M. M.; Suntzeff, N. B.; Clocchiatti, A.; Piemonte, A.; Leibundgut, B.; Covarrubias, R.; Maza, J.; Sollerman, J.,
%1998, ApJ, 498, L129

\bibitem[\protect\citeauthoryear{Utrobin}{ 2007}]{Utrobin-99emref}
Utrobin V. P., 2007, A\&A, 461, 233U

\bibitem[\protect\citeauthoryear{Utrobin \& Chugai}{ 2008}]{Utrobin-05csref}
Utrobin V. P., Chugai N. N., 2008, A\&A, 491, 507

\bibitem[\protect\citeauthoryear{Utrobin \& Chugai}{ 2009}]{Utrobin-04etref}
Utrobin V. P., Chugai N. N., 2009, A\&A, 506, 829U

\bibitem[\protect\citeauthoryear{Van Dyk et al.}{ 2012}]{Vandyk-12awrefHST}
Van Dyk S. D. et al., 2012, ApJ, 756, 131V
%Van Dyk, Schuyler D.; Cenko, S. Bradley; Poznanski, Dovi; Arcavi, Iair; Gal-Yam, Avishay; Filippenko, Alexei V.; Silverio, Kathryn; 
%Stockton, Alan; Cuillandre, Jean-Charles; Marcy, Geoffrey W.; Howard, Andrew W.; Isaacson, Howard, 2012, ApJ, 756, 131V

\bibitem[\protect\citeauthoryear{Wanajo et al.}{ 2008}]{Wanajo-ecapture}
Wanajo S., Nomoto K., Janka H.-Th., Kitaura F. S., Mueller B., 2009 ApJ 695, 208

%pozitron felszabadulas
\bibitem[\protect\citeauthoryear{Woosley et al.}{ 1989}]{Woosley-leak}
Woosley S. E., Hartmann D., Pinto P. A., 1989, ApJ, 346, 395W

%\bibitem[\protect\citeauthoryear{SDSS DR15}{}]{sdss-dr9} %%%%%%%%%%%%%%%%%%%%%%%%%%%%%%%%%%%%!!!!!!!!!!!!!!!!!!!!!!!!!
%sdss catalogs data release 15, http://skyserver.sdss.org/dr9/en/tools/chart/navi.asp



\end{thebibliography}
\end{document}